\documentclass[twocolumn]{aastex62}
\usepackage{subfigure} 
\pdfoutput=1 
\usepackage{amsmath,amstext}
\usepackage[utf8]{inputenc}
\usepackage{apjfonts} 
\usepackage[figure,figure*]{hypcap}
\usepackage[inline]{enumitem}   
\usepackage{longtable}
\usepackage{tabularx}

\shorttitle{WEIRD}
\shortauthors{Baron et al.}
 
\newcommand{\mj}{\,$M_{\rm Jup}$}

\begin{document}

\title{WEIRD: Wide-orbit Exoplanet search with InfraRed Direct imaging}
 
\author{Frédérique Baron}
\affiliation{Institut de Recherche sur les Exoplanètes, Département de Physique, Université de Montréal, Montréal, QC H3C 3J7, Canada}
\author{Étienne Artigau}
\affiliation{Institut de Recherche sur les Exoplanètes, Département de Physique, Université de Montréal, Montréal, QC H3C 3J7, Canada}
\author{Julien Rameau}
\affiliation{Institut de Recherche sur les Exoplanètes, Département de Physique, Université de Montréal, Montréal, QC H3C 3J7, Canada}
\author{David Lafrenière}
\affiliation{Institut de Recherche sur les Exoplanètes, Département de Physique, Université de Montréal, Montréal, QC H3C 3J7, Canada}
\author{Jonathan Gagné}
\affiliation{Carnegie Institution of Washington DTM, 5241 Broad Branch Road NW, Washington, DC 20015, USA }
\affiliation{NASA Sagan Fellow}
\author{Lison Malo}
\affiliation{Institut de Recherche sur les Exoplanètes, Département de Physique, Université de Montréal, Montréal, QC H3C 3J7, Canada}
\author{Loïc Albert}
\affiliation{Institut de Recherche sur les Exoplanètes, Département de Physique, Université de Montréal, Montréal, QC H3C 3J7, Canada}
\author{Marie-Eve Naud}
\affiliation{Institut de Recherche sur les Exoplanètes, Département de Physique, Université de Montréal, Montréal, QC H3C 3J7, Canada}
\author{René Doyon}
\affiliation{Institut de Recherche sur les Exoplanètes, Département de Physique, Université de Montréal, Montréal, QC H3C 3J7, Canada}
\author{Markus Janson}
\affiliation{Department of Astronomy, Stockholm University, SE-106 91 Stockholm, Sweden}
\author{Philippe Delorme}
\affiliation{Université de Grenoble Alpes, CNRS, IPAG, F-38000 Grenoble, France}
\author{Charles Beichman}
\affiliation{NASA Exoplanet Science Institute, California Institute of Technology, Jet Propulsion Laboratory, Pasadena, CA 91125, USA}

\begin{abstract}

We report results from the Wide-orbit Exoplanet search with InfraRed Direct imaging (WEIRD), a survey designed to search for Jupiter-like companions on very wide orbits (1000 to 5000~AU) around young stars ($<$120~Myr) that are known members of moving groups in the solar neighborhood ($<$70~pc). Sharing the same age, distance, and metallicity as their host while being on large enough orbits to be studied as "isolated" objects make such companions prime targets for spectroscopic observations and valuable benchmark objects for exoplanet atmosphere models. The search strategy is based on deep imaging in multiple bands across the near-infrared domain. For all 177 objects of our sample, $z_{ab}^\prime$, $J$, [3.6] and [4.5] images were obtained with CFHT/MegaCam, GEMINI/GMOS, CFHT/WIRCam, GEMINI/Flamingos-2, and $Spitzer$/IRAC. Using this set of 4 images per target, we searched for sources with red $z_{ab}^\prime$ and $[3.6]-[4.5]$ colors, typically reaching good completeness down to 2~\mj\ companions, while going down to 1~\mj\ for some targets, at separations of $1000-5000$~AU. The search yielded 4 candidate companions with the expected colors, but they were all rejected through follow-up proper motion observations. Our results constrain the occurrence of 1--13~\mj\ planetary-mass
companions on orbits with a semi-major axis between 1000 and 5000~AU at less than 0.03, with a 95\% confidence level. 

\end{abstract}
 

\section{Introduction} 
Since the first detection of an exoplanet around a main sequence star by \cite{mayor_jupiter-mass_1995}, thousands of exoplanets have been discovered, revealing planetary system architectures vastly different from that of the Solar System. The most successful methods to detect exoplanets are the transit and radial velocity methods, which are more effective for planets close to their host star (up to 15 AU). The outer part of planetary systems can be probed by direct imaging. The first detection of a giant planet by direct imaging was made in 2004, with the discovery of a 4~\mj\ planet orbiting the brown dwarf 2MASSW~J1207334-393254 \citep{chauvin_giant_2004}, and the search for directly imaged planets has continued since then.

A good number of direct imaging surveys for planetary mass objects on wide obits were carried out in the last decade. Some targeted only low mass stars, such as \cite{bowler_planets_2015}, \cite{lannier_massive:_2016}, and \cite{naud_psym-wide:_2017}, while others surveyed higher mass stars \citep{vigan_international_2012,nielsen_gemini_2013,rameau_survey_2013} or all spectral types \citep{lafreniere_gemini_2007,heinze_constraints_2010,biller_gemini/nici_2013,chauvin_vlt/naco_2015}. \cite{bowler_imaging_2016} did a meta-analysis using data from the most complete studies that surveyed all types of star \citep{lafreniere_gemini_2007,janson_high-contrast_2011,vigan_international_2012,biller_gemini/nici_2013,janson_seeds_2013,nielsen_gemini_2013,wahhaj_gemini_2013,brandt_statistical_2014,bowler_planets_2015} using 384 stars with spectral types B2 to M6. He obtained an overall planet occurrence rate for BA, FGK, and M stars of respectively 2.8$_{-2.3}^{3.7}$ \%, $<$4.1\%, and $<$3.9\% for 5-13~\mj\ planets at separations of 30 to 300 AU.

Direct imaging surveys have typically targeted young stars, which are prime targets since their planets are still contracting and are thus warmer and brighter than their older counterparts, for a given mass. The number of known young stars near the Sun has dramatically increased in the last few years, as a result of a growing interest for young stellar moving groups \citep[e.g.,][]{zuckerman_young_2004,torres_young_2008}. A moving group is composed of stars that were formed together  less than a few hundreds of Myr ago, and therefore still share similar $UVW$ galactic velocities, enabling their identification. In the recent years, a significant effort has been made to identify a large number of low mass stars, brown dwarfs and isolated planetary-mass objects that are members of known young moving group \citep{lepine_nearby_2009,shkolnik_identifying_2009,kiss_search_2011,schlieder__2010,rodriguez_new_2011,liu_extremely_2013,schlieder_cool_2012,schlieder_likely_2012,shkolnik_identifying_2012,malo_bayesian_2013,moor_unveiling_2013,rodriguez_galex_2013,kraus_stellar_2014,malo_banyan._2014,riedel_solar_2014,gagne_banyan._2014,binks_kinematically_2015,gagne_banyan._2015}.

Planets found on wide orbits around young stars are very interesting because they can be characterized much better than their closer-in counterparts. First, a planet bound to a well-studied star shares some properties with it, like its age, distance from Earth, and metallicity. Furthermore,  when a planet is on an large enough orbit, it can be studied as if it were an isolated  object, that is without adaptive optics \citep{naud_discovery_2014,gauza_discovery_2015}, and a very high resolution spectrum can then be acquired, which is very hard to obtain for closer-in planets. Also, the large separation to the host enables direct studies that are very challenging with high-contrast imaging (e.g., accurate spectro-photometry, intermediate resolution spectroscopy, optical imaging, time-variability). Such planetary mass objects are also prime targets for $JWST$ follow-up.

Widely separated systems are of further interest because they challenge formation processes. Theories predict that giant planets form either by core accretion or gravitational instability, or like brown dwarfs by cloud fragmentation. The former process describes a way of forming planets by first building a 5 to 20  $M_{\rm Earth}$ core of rocks and ices, in a protoplanetary disk \citep{alibert_planet_2009,inaba_formation_2003,pollack_formation_1996}. The core then accretes gas very rapidly to form a giant planet. This method explains very well the formation of planets on close-in orbit \citep[$<$ 10\,AU,][]{mordasini_characterization_2012}, but struggles to explain the formation of planets on wide orbits. The second process suggests that planets form from the fragmentation of a gravitationally unstable disk \citep{boss_formation_2011}, which forms clumps that then can accrete gas and dust to become planets \citep{stamatellos_brown_2007,bate_stellar_2012}. However, this mechanism also has difficulties forming planets on wide orbits, as shown for example by \cite{nayakshin_desert_2017} and \cite{vigan_vlt/naco_2017}. 
The last process predicts that planets on wide orbits form from the direct collapse of the molecular cloud \citep{padoan_``mysterious_2004}. A fragment of a few Jupiter mass is formed which then accretes gas from the cloud to form a higher mass object. However, \cite{bate_formation_2002} and \cite{bate_stellar_2012} have shown that the accreation process can be stopped at a low mass if the companion is ejected away from the dense part of the envelope or if the envelope is depleted at the formation time. However this formation process tends to form preferentially equal mass binaries and does not seem to produce systems with the high mass ratios needed to match the observed planetary systems at wide separations.
Dynamical instabilities are a tantalizing alternative to explain the detected planets at large separations \citep{chatterjee_dynamical_2008,veras_formation_2009,baruteau_recent_2013}. Mutual gravitational perturbations and close encounters among the planets occur and pump the eccentricity and the semi-major axis of the less massive giant planets up to 100 - 100 000\,AU \citep{veras_formation_2009}, but close-in scatterers  are yet to be discovered \citep{bryan_searching_2016}.

We report here the results from the Wide-orbit Exoplanet search with InfraRed Direct imaging (WEIRD). The WEIRD survey started in 2014 with the aim to detect Jupiter-like companions on very wide orbits (at separations 1000--5000\,AU) around all the known members of young moving groups within 70~pc. We gathered a large dataset 
to try to construct the SED of such objects through
of deep $[3.6]$ and $[4.5]$ imaging from $Spitzer$/IRAC and deep seeing-limited $J$ and $z_{ab}^\prime$ imaging from CFHT and Gemini-South of all 177 known (at the time) young (<120 Myr) objects in a volume-limited sample of 70~pc of the Sun. Using these data, planetary companions can be revealed through their distinctively red $z_{ab}^\prime-J$ and $[3.6]-[4.5]$ colors. The range of separations studied here has been barely probed by previous direct imaging surveys as they were limited by the field of view of high contrast imagers, with the exception of \cite{naud_psym-wide:_2017}, which was much less sensitive than the present survey, and limited to low-mass stars. The selection of the sample of young stars and the observing strategy and data reduction are described in Section~\ref{sample}. Section~\ref{results} presents the results of our search, while Section~\ref{discussion} discusses the statistical analysis of the survey.

\begin{table*}
\center
\caption{Young Moving Groups\label{asso}}
\begin{tabular}{ccccc}
\hline\hline Name& Distance& Age & \textbf{Number of members detected} & Ref. \\ &(pc)&(Myr)&\\\hline
N$\beta$-Pictoris & 9-73 & 24$\pm$3 &51& \cite{shkolnik_all-sky_2017} \\\hline
AB-Doradus & 37-77 & 149$^{+51}_{-19}$ &58& \cite{bell_isochronal_2016}\\\hline
Argus &  8-68 & 30-50 &10& \cite{torres_young_2008}\\\hline
Carina & 46-88 & 45$^{+11}_{-7}$ &6&\cite{bell_isochronal_2016} \\\hline
Columba &  35-81 & $42^{+6}_{-4}$&15&\cite{bell_isochronal_2016}  \\\hline
Tucana-Horologium& 36-71 & 45$\pm$3 & 50& \cite{bell_isochronal_2016} \\\hline
TW Hydrae& 28-92  & 10$\pm$3 &16& \cite{bell_isochronal_2016}\\\hline
\end{tabular}
\end{table*}

\section{Sample and Observations}\label{sample}
\subsection{Sample}
The best targets to find giant planets on very wide orbits are young stars in the solar neighborhood because giant planets are more luminous when they are younger and they become fainter with time. Therefore, observations of  younger stars are sensitive to lower-mass planets. 
A sample was thus created by selecting all stars within 70~pc that are members of the following young moving groups or associations (see Table~\ref{asso}):
TW Hydrae \citep{de_la_reza_discovery_1989,kastner_x-ray_1997},
$\beta$ Pictoris \citep{zuckerman__2001},
AB Doradus \citep{zuckerman_ab_2004},
Tucana Horologium \citep{torres_new_2000,zuckerman_tucana_2001},
Carina \citep{torres_young_2008},
Columba \citep{torres_young_2008}
and Argus \citep{makarov_moving_2000}.
The members of these groups have ages in the range 10–150\,Myr. The age of the Argus moving group is not well constrained, likely because current membership lists suffer from significant contamination from unrelated field-aged stars \citep{bell_isochronal_2016}. To be considered bona fide members of one group and included in our sample, the stars must have a trigonometric parallax and a radial velocity measurement, $XYZUVW$ values consistent with the moving group membership, as well as independent signatures of youth, e.g. spectroscopic signs of low-gravity, strong X-ray or UV emission or lithium absorption \citep[see][]{soderblom_ages_2010}. The sample was constructed from \cite{gagne_banyan._2014}; \cite{kiss_search_2011};  \cite{lepine_nearby_2009}; \cite{malo_bayesian_2013}; \cite{schlieder__2010}
; \cite{shkolnik_identifying_2009} ; \cite{shkolnik_searching_2011} ; \cite{shkolnik_identifying_2012} ; \cite{song_new_2003}; \cite{torres_new_2000} ; \cite{torres_young_2008}; \cite{zuckerman_identification_2000}; \cite{zuckerman_ab_2004}; \cite{zuckerman__2001} ; \cite{zuckerman_tucana_2001-1} ; \cite{zuckerman_tucana/horologium_2011}. We note that these publications also proposed a larger sample of strong candidates but they lacked one or more measurements to be confirmed members; these objects were not included in our sample. Our complete sample includes 177 objects.

Multiple systems were not excluded from the sample as the presence of a lower or equal mass object does not exclude the possibility of having a planetary mass object on a wide orbit. For example, Ross 458 (AB)c is a triple system comprising a tight M0.5+M7 binary orbited by an 11~\mj\ object \citep{goldman_new_2010} and 2MASS J01033563-5515561(AB)b, a 12--14 \mj\ object, orbits a pair of young late-M stars at 84 AU\citep{delorme_direct-imaging_2013}. Also, 
\cite{wang_influence_2015} have shown that stellar multiplicity does not influence the presence of planets on wide (100 to 2000\,AU) orbits in the system. In our sample of targets, 68 are multiple systems, 2 host brown dwarf companions \citep{schneider_nicmos_2004,chauvin_companion_2005} and 4 host known planets \citep{chauvin_giant_2004,marois_direct_2008,lagrange_probable_2009,macintosh_discovery_2015}.

\begin{figure*}[t]
\centering
\plottwo{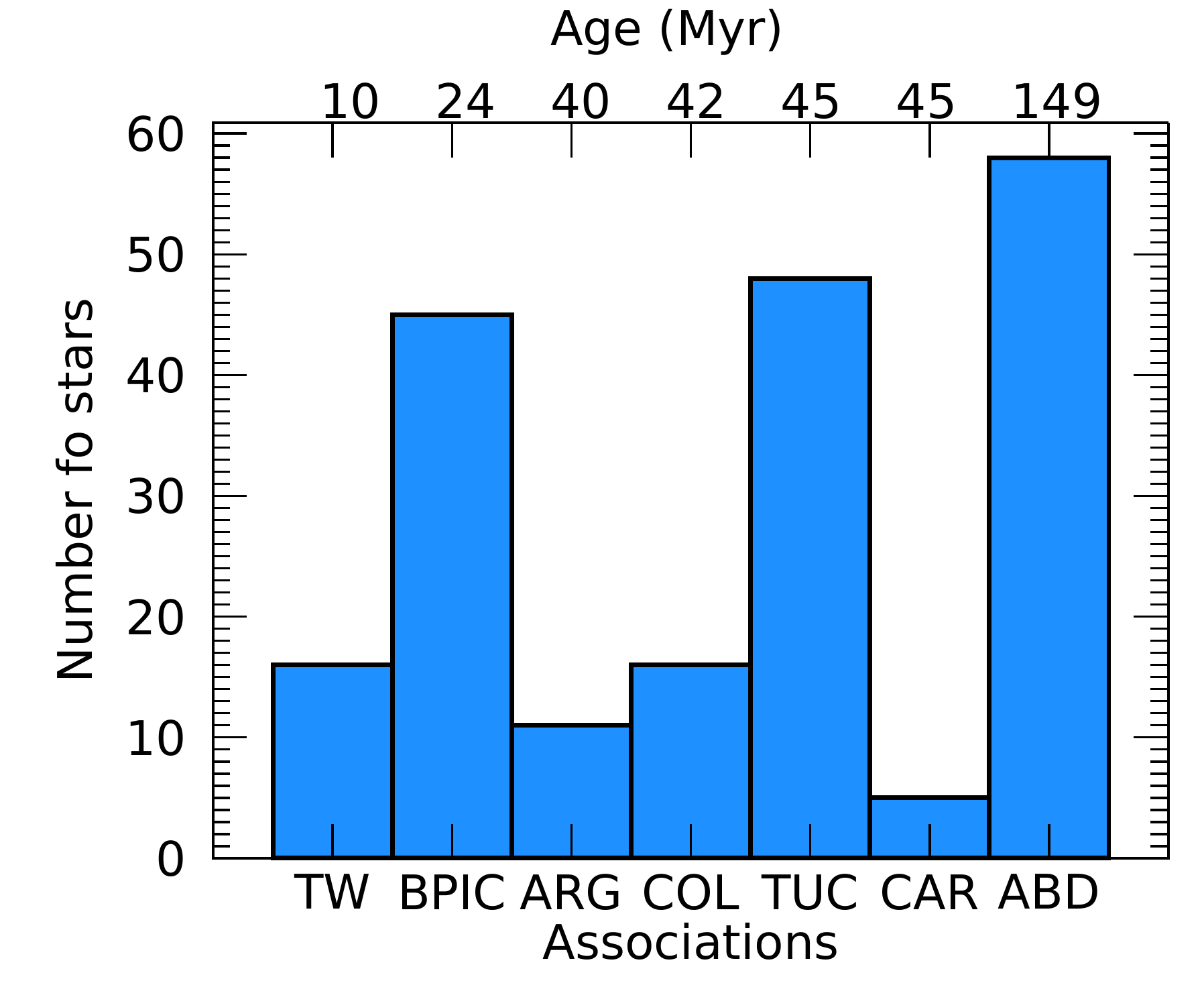}{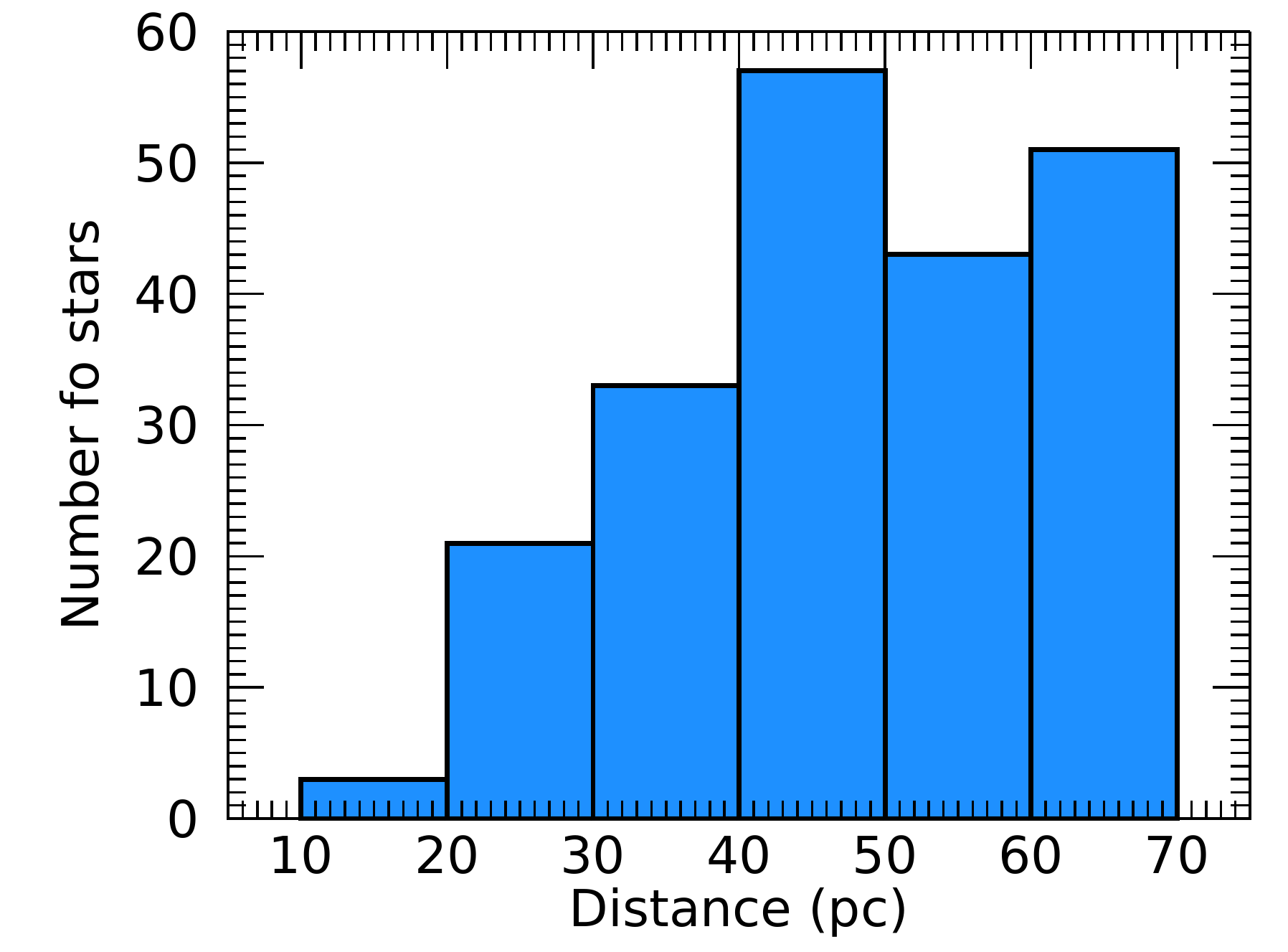}
\plottwo{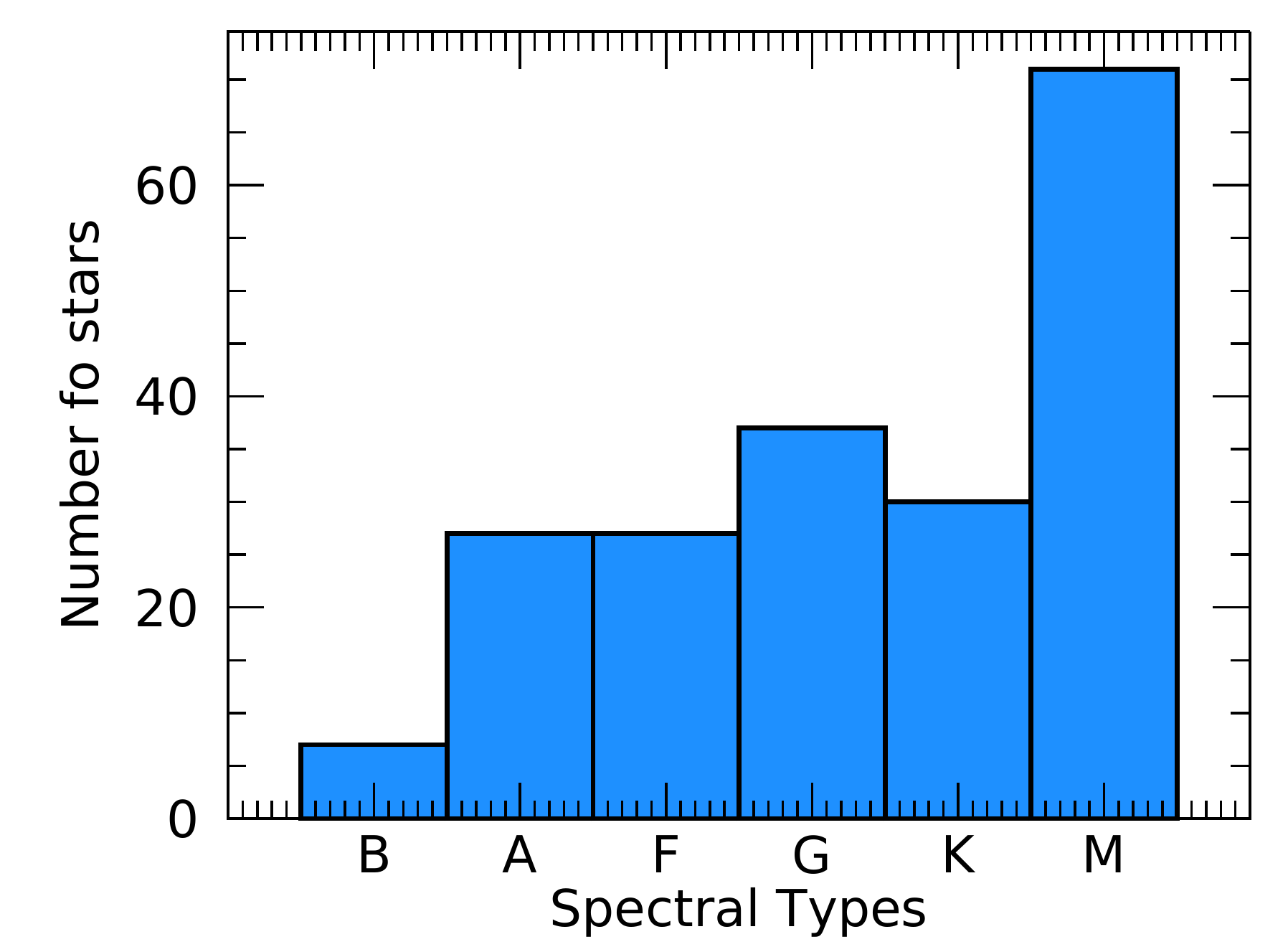}{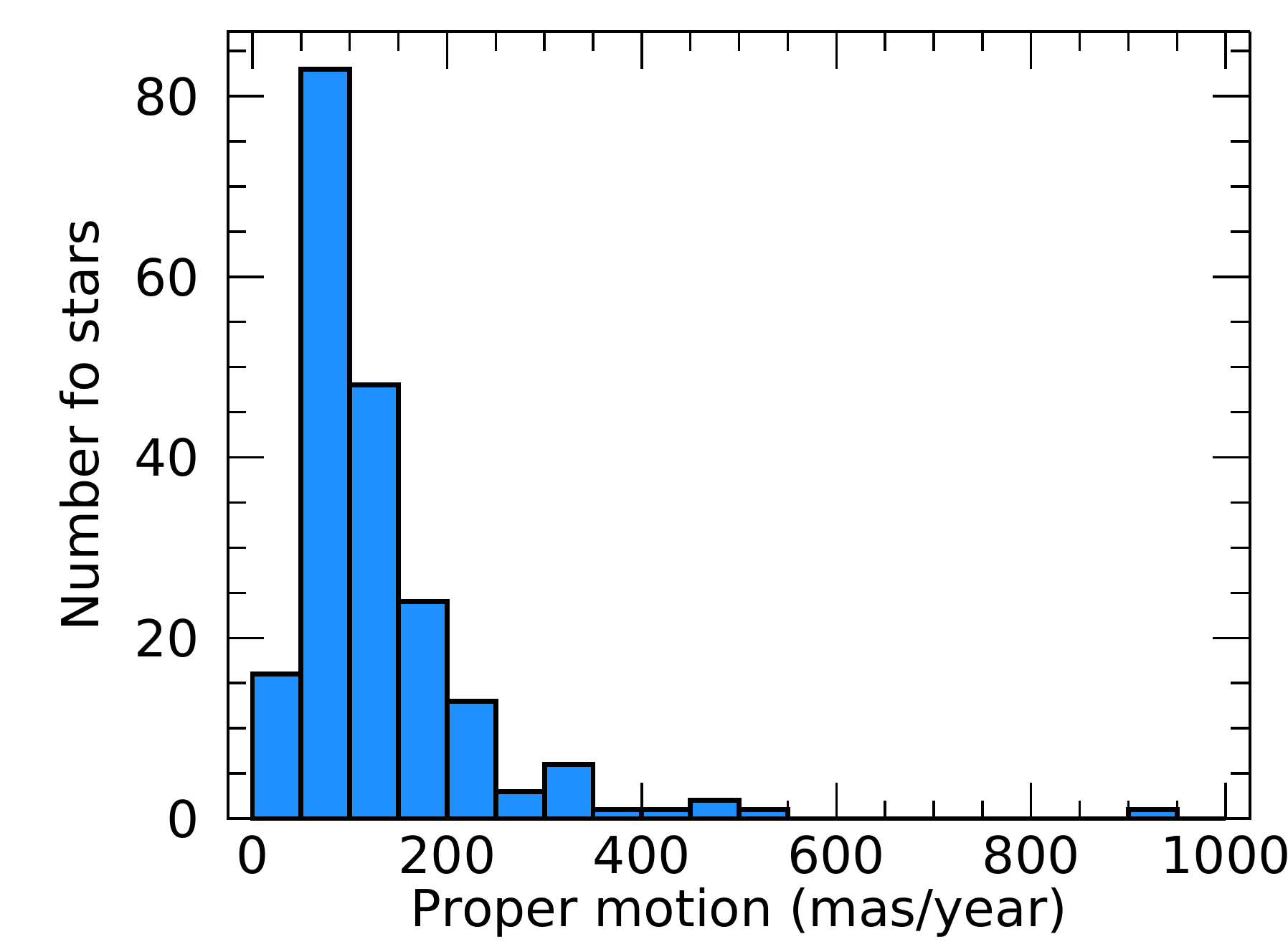}

\caption{\label{figcarac} Distributions of associations, distances (pc), spectral types and proper motions (mas/yr) for all the stars in the sample.}
 	
 \end{figure*}

The properties of the 177 objects in our sample is presented in Table~\ref{listecomplete}. They have spectral classes in the range A--M, with a majority of M dwarfs, are located at distances of 7 to 70 pc, are located all over the sky, and have relatively high proper motions (see Figure~\ref{figcarac}). The median star has a proper motion of 100 mas/yr, a distance of 42 pc and an age of 45 Myr. Table~\ref{listeasso} lists all the systems in our sample along with their radial velocities, distances and the association they are a member of. 

\subsection{Observing strategy}

\begin{figure*}[ht] 
\centering
\plottwo{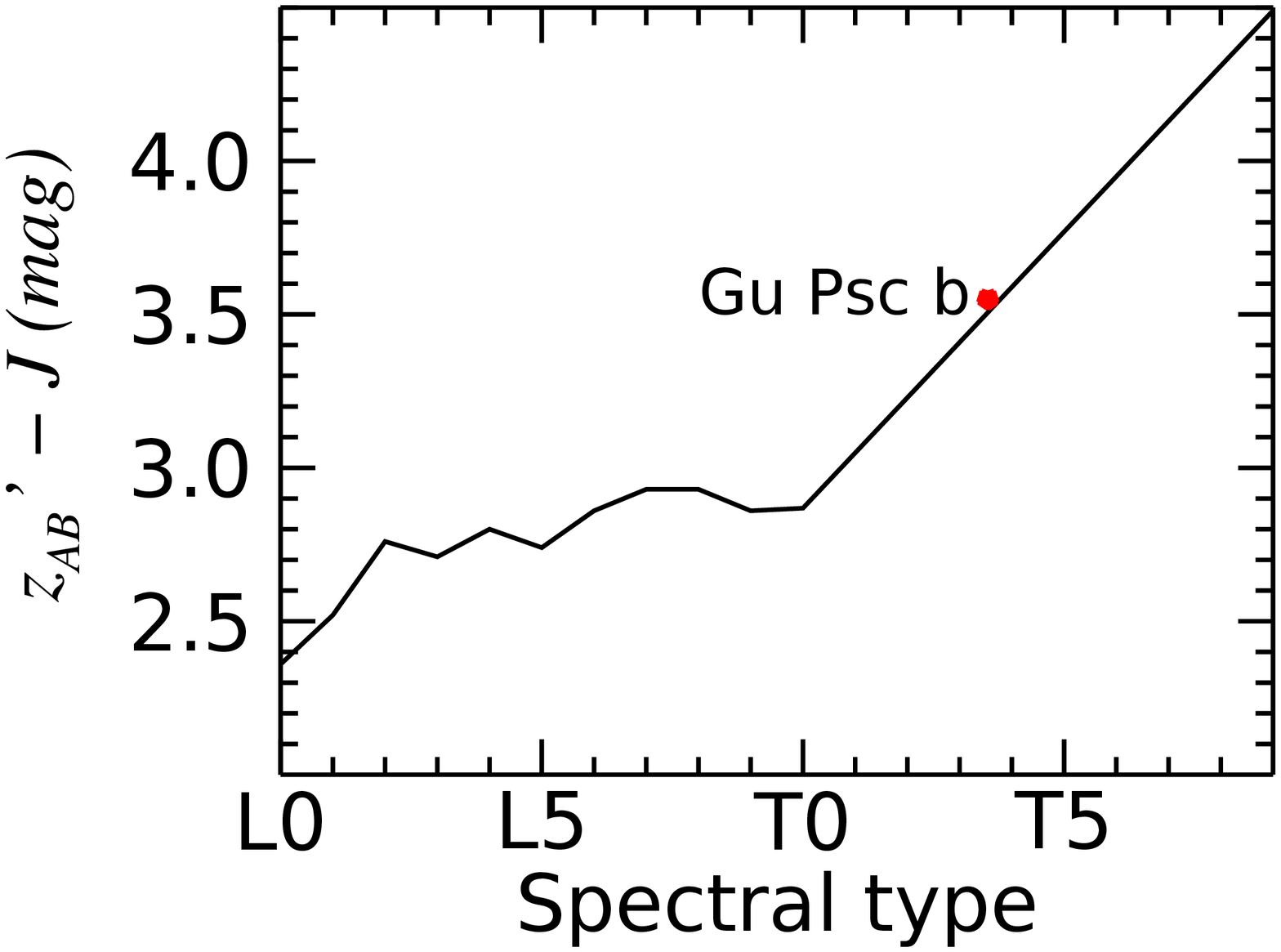}{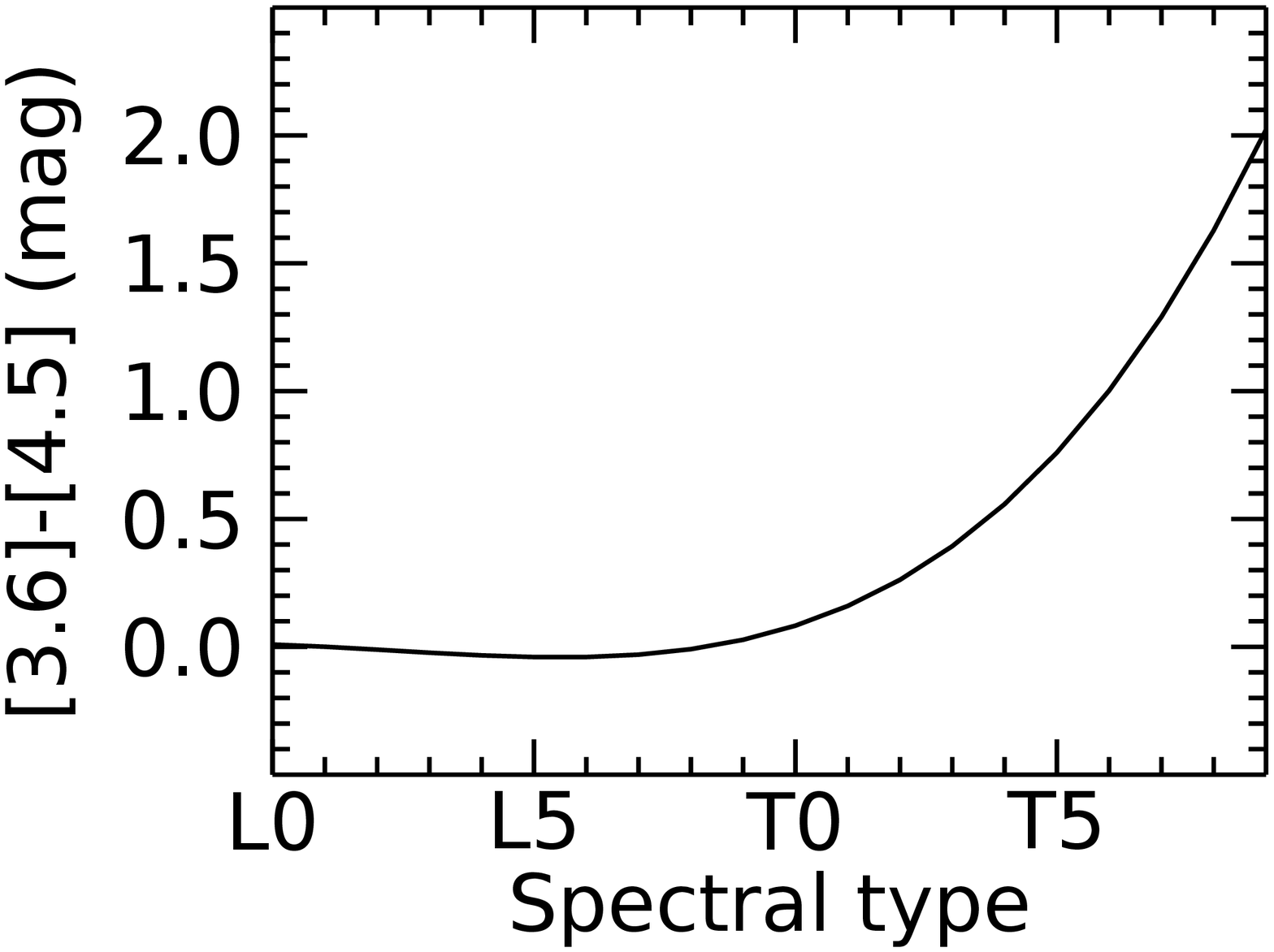}
\caption{\label{col_spt} On the left, $z_{ab}^\prime-J$ vs spectral type for L to T dwarfs from \cite{hawley_characterization_2002} for the L dwarfs and \cite{albert_37_2011} for the T dwarfs. The L to T dwarfs are characterized by red $z_{ab}^\prime-J$ colors. The red dot represents Gu Psc b, the planetary mass object discovered by \cite{naud_discovery_2014}, representative of the kind of objects we are seeking in this work. On the right,  [3.6]-[4.5] for L to T dwarf from \cite{dupuy_hawaii_2012}. We see that late T dwarfs can be identified both by their red [3.6]-[4.5] $>$ 1.5 and $z_{ab}^\prime-J$ $>$ 4 colors.  }
 \end{figure*}

\begin{figure*}[t]
\centering
\epsscale{1} 
\plotone{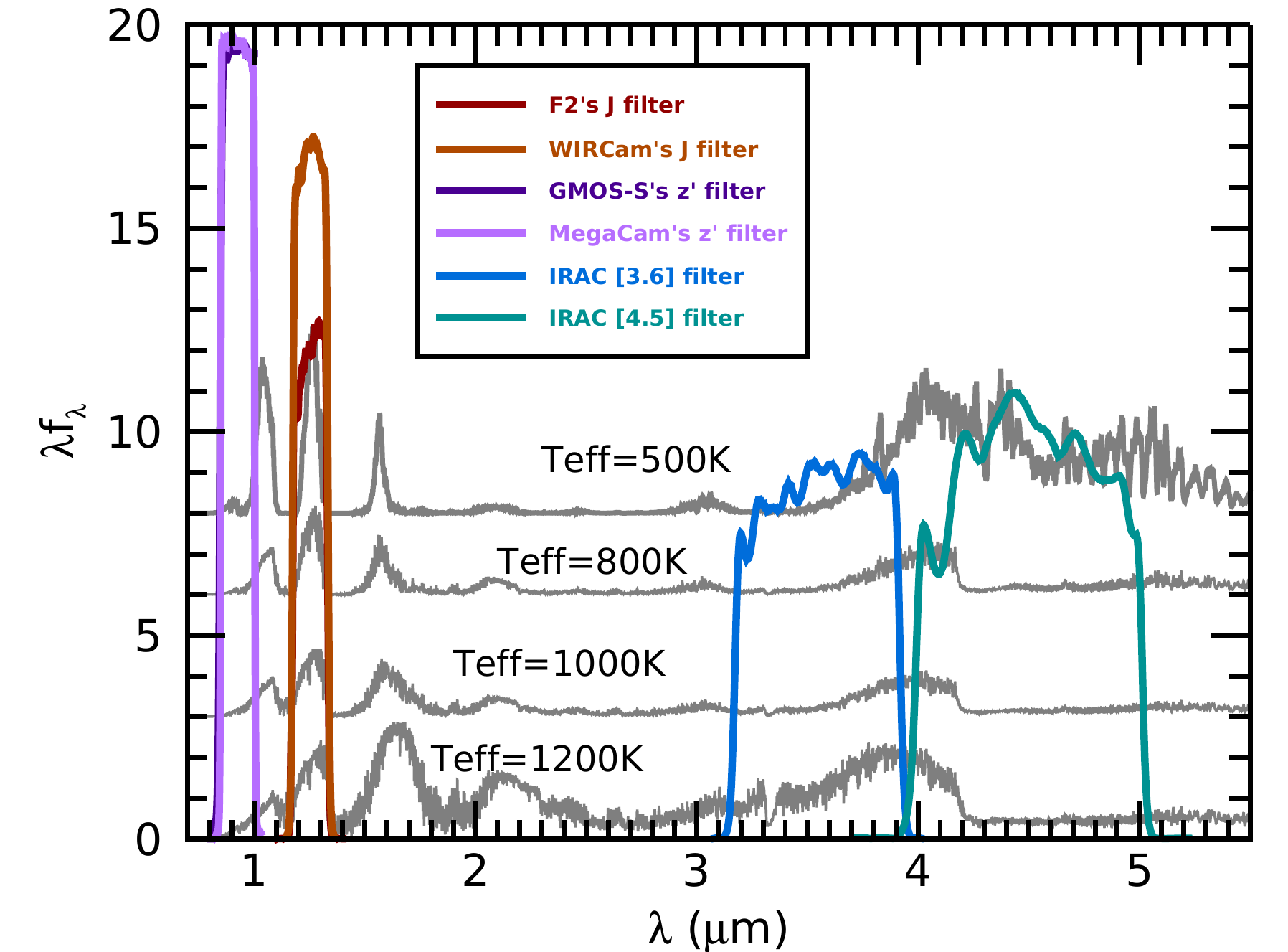}
\caption{\label{filters}
BT-Settl spectral energy distribution of young objects ($log g$ =4 and solar metallicity) with effective temperatures of 500~K, 800~K, 1000~K, and 1200~K. The transmission functions of the four bandpasses used for our observations ($z_{ab}^\prime$, $J$, $[3.6]$, and $[4.5]$) are overlaid. These bandpasses provide distinctive red colors while maintaining a high flux level across the temperature range.
}
\end{figure*}

Figure~\ref{col_spt} presents the typical $z_{ab}^\prime - J$ and $[3.6]-[4.5]$ colors as a function of spectral type for objects ranging from spectral types L to T, a range relevant for the companions sought here. It shows that both L and T dwarfs have very red $z_{ab}^\prime-J$ colors, with the color of an L dwarf being between 2.5 and 3 mag, and the color of a T dwarf between 3 and 4.5 mag. Beyond those types, as shown by \cite{lodieu_gtc_2013}, the $z_{ab}^\prime - J$ colors of Y dwarfs remain red but vary much more, ranging from from 2.5 to 5 mag. In the mid-infrared, starting at around T0, the $[3.6]-[4.5]$ color becomes increasingly red with spectral type, reaching values larger than 1.5~mag for late-T's. Young objects, with larger radii and correspondingly lower surface gravities, would have slightly redder colors compared to the colors of field dwarfs shown in the figure.
The strategy used in the current survey builds on these markedly red colors of very late-type dwarfs across these four spectral bands, which enables distinguishing them easily from earlier-type objects and most other astrophysical sources. In addition, as shown in Figure~\ref{filters}, these bands are also optimal to maximize the flux of the objects sought over the temperature range of interest.

The ground-based component of our survey is optimized to find companions up to spectral type $\sim$T9, while the Spitzer component is optimized for later types. At a distance of 42 pc (the median distance of our sample), the expected $J$ magnitude of a T9 dwarf is about 21~mag. We thus designed our observations in $J$-band to reach 21~mag. As T dwarfs later than T0 are expected to have $z_{ab}^\prime-J$ > 3~mag, we designed our observations to reach $z_{ab}^\prime=24$~mag as they can be identified either through detection in both bands or as $z_{ab}^\prime$ dropouts. For the this part of the survey, we used the same detection criteria as for the CFBDSIR survey \citep{delorme_cfbds_2008,delorme_extending_2010,albert_37_2011}. That survey was a wide-field search for T dwarfs and early-type Y dwarfs, and the candidates were identified through their very-red $z_{ab}^\prime-J$ > 3~mag colors if they were detected in both bands, or through $z_{ab}^\prime$dropouts.  The CFBDSIR survey returned only 64 candidates over the 280 square degrees observed, of which 17 were actual field T dwarfs. The strategy of searching for very low-mass objects using NIR colors has also been employed by the PSYM-wide survey \citep{naud_psym-wide:_2017} to probe nearby young M dwarfs for planetary mass companions. The survey discovered a new planetary mass object (9–13\,$M_{\rm Jup}$) orbiting at 2000\,AU around the M3V star Gu Psc, a highly probable member of the AB Doradus moving group \citep{naud_discovery_2014}\footnote{Using the parallax of $21.0019\pm0.0721$~mas \citep{lindegren_gaia_2018,gaia_collaboration_gaia_2018} for GU Psc from the \textit{Gaia} DR2 release, and the web tool BANYAN $\Sigma$ from \cite{gagne_banyan._2018}, we infer that the probability of Gu Psc to be a member of the AB Doradus moving group is 99.1 \%, which confirms the membership of the star.} .

The $Spitzer$/IRAC observations were designed so that they reach a sufficient depth to identify point-sources in the field down to $\sim$ $0.5$\,mag of the confusion limit and have their color measured with accuracy to identify them at $>5\sigma$ level compared to the bulk of background objects. We perform the point-source detection in $[3.6]$, which provides deeper images for flat-spectrum sources, and use the $[4.5]$ photometry to constrain colors. Our observations are designed to reach depths of 21.2~mag ($5\sigma$) and 20.7~mag ($3\sigma$) in $[3.6]$ and $[4.5]$, respectively. Planetary-mass objects close to the detection limit, with masses below 3-5~\mj\, will have $[3.6]-[4.5]>2$~mag (see right-hand side Figure~\ref{col_spt}) and will therefore be detected at a higher signal-to-noise ratio in $[4.5]$. Because they would have M$_J> 18$~mag (or $J>21$~mag for a typical target) or a spectral type $\gtrsim$T8.5, such objects would be $z_{ab}^\prime$- and $J$-band dropouts ($z_{ab}^\prime- [4.5] > 6$~mag and $J-[4.5] > 3.5$~mag). Given that background objects typically have $[3.6]-[4.5]$$\sim0.0 \pm0.4$~\,mag, such planetary-mass objects will differ from the bulk of background objects at the $>$$5\sigma$-level. However, by itself this part of the dataset is insensitive to more massive (> 3-5~\mj) companions as their colors don't differ enough from those of background objects.

\subsection{Observations and Data Reduction} \label{obs}

All targets in our sample were observed with deep seeing-limited $J$ and $z_{ab}^\prime$ imaging at either the Canada-France-Hawaii Telescope (CFHT) with WIRCam \citep{puget_wircam:_2004} and Megacam \citep{boulade_megacam:_1998}, or at Gemini-South with GMOS-S \citep{hook_gemini-north_2004,gimeno_-sky_2016} and Flamingos-2 \citep{eikenberry_flamingos-2:_2012}, as well as with $Spitzer$/IRAC \citep{fazio_infrared_2004} in the $[3.6]$ and $[4.5]$ bands. Stars with a declination $< - 35 \degr$ were observed from the ground at the Gemini-South Observatory while the others were observed at the CFHT. Throughout this work, all the $J$-band magnitudes are in the Vega system while all the $z_{ab}^\prime$ magnitudes are in the AB system. For a median star in our sample with a distance of about 42~pc and an age of 45~Myr, the limiting magnitude in both bands corresponds to $M_{J}=17.9$~mag and $M_{z}=20.9$~mag, or to an effective temperature of about 385~K according to models from \cite{baraffe_evolutionary_2003}.

\subsubsection{Gemini Observations}

The observations were made from 2014 to 2017 at Gemini-South (GS-2014B-Q-2, GS-2015A-Q-71, GS-2015B-Q-57, GS-2016A-Q-69, GS-2016B-Q-33, GS-2017A-Q-58, PI Frederique Baron). We obtained deep imaging of our southern sub-sample with Flamingos-2 with the $J$ filter (J\_G0802, 1.255~$\mu$m) and the Gemini Multi-Object Spectrograph (GMOS) in the $z_{ab}^\prime$ filter (z\_G0328, $>$848 $\mu$m).  Objects beyond 30 pc, the vast majority of our targets, are sufficiently far for the entire projected 5000~AU sphere around them to fit within the GMOS/F2 FOV.

Flamingos-2 is a near-infrared wide-field imager and multi-object spectrometer with a 6.19 arcmin$^2$ circular field of view and a 0.18\arcsec pixel scale. We obtained at least 600s of integration time on each target, divided into a different number of expositions (at least 9) depending on the magnitude of the star and the observing conditions. A small random dither pattern was used to mitigate detector defects. The exposition time was selected to reach a limiting magnitude of $J=21$~mag at a $7\sigma$ level. Each observation was about 20 minutes long, including all overheads.  

GMOS has three 2048$\times$4608 CCDs which, when combined, have a field of view of 5.5x5.5 arcmin$^2$ and a pixel scale of 0.073\arcsec. We obtained 8 expositions of 65 s for each target of the sample. A dither pattern of 17 \arcsec was used for all observations.  The exposure time was chosen to reach a limiting magnitude of $z=24$~mag at $3\sigma$. The observations were each about 20 minutes long, including all overheads.

The $J$-band images from F2 were reduced using a custom IDL pipeline. The individual images were reduced by subtracting dark images, dividing by flat field images, and correcting the residual gradient from the vignetting of the  Peripheral Wavefront Sensor (PWFS). This step was done by first  normalizing the image to its median value, then masking regions with values significantly over the median to get rid of the stars. This image was in turn used to create a gradient image where each pixel is the median of a 128x128 pixel box of the masked image. A polynomial fit of degree 3 was then applied to the gradient image. This was divided from the F2 image to correct for the vignetting by the PWFS. The astrometric correction was then computed by anchoring the star positions on the \emph{Gaia}~DR1 catalog \citep{gaia_collaboration_gaia_2016}. A radial profile about the bright target star was then subtracted to help search for sources at smaller separations.  Lastly, a low-pass filter was created by median binning the image by 4x4 pixels, applying a 15x15-pixel median filter, and then resampling at the original image size. This low-pass filter was subtracted from the image to facilitate the detection of point sources. The individual images for a given target were then combined by taking their median, after astrometric registration, to produce the final $J$-band image.

For GMOS, the images were also reduced using a custom IDL reduction pipeline. Each CCD was first processed separately. First, a sky correction was applied by subtracting the median of all images taken on a given night. When needed, any detector region affected by the on-instrument wave front sensor was masked. Most of the time the wave front sensor was off the detector, but sometimes it was not possible to find a guide star outside of the FOV. The astrometric solution was found for each CCD by anchoring the field to the F2 reduced image of the same target. A high-pass filter was then applied by subtracting a median-filtered image with a width of 15 pixels. A one-dimensional median-filter with a width of 61 pixels was also subtracted from each line of the image to correct for the saturation banding. The 3 CCDs were then combined to form a complete image, to which the astrometric solution was applied again. All the images for a given target were then aligned and stacked by taking their median to get the resulting reduced image.

\subsubsection{CFHT Observations}
Deep imaging of our northern sub-sample was obtained at the CFHT from 2014 to 2017  using WIRCam with the $J$ filter and MegaCam with the $z_{ab}^\prime$ filter (14BC016, 15AC032, 15BC012, 16AC021, 16BC018, 17AC23; PI Frédérique Baron).

WIRCam \citep{puget_wircam:_2004} is a near infrared wide-field imager with a field-of-view of 20 arcmin$^2$ and a pixel scale of 0.3\arcsec. It uses a mosaic of 4 detectors with a small gap between each. We used the $J$-band (1.253 $\mu$m) filter and a homemade dither pattern of 16 60-s expositions, arranged so that the target does a small dither of 28\arcsec around a pixel situated 64\arcsec from the corner of one detector near the center of the field, for a total of 1120~s of on-target integration time. A different dither pattern was used to mitigate the saturation effects of stars brighter than $J=7$. In those cases, the bright target was put in a gap between quadrants at each position of the dither pattern. With a seeing between 1\arcsec and 1.2\arcsec, the exposure time is sufficient to reach a SNR=7 at a limiting magnitude of $J\approx 21$.

MegaCam \citep{boulade_megacam:_1998} is a wide-field optical imager with a 1 square degree field-of-view and a pixel scale of 0.187\arcsec. We used the $z_{ab}^\prime$ filter (z\_G0328, $>$848 $\mu$m) and a dither pattern with 4 positions offset by 15\arcsec, which is twice the size of the standard dither pattern. The total integration time per target varies between 311~s and 476~s. The higher integration time is for targets with a declination in the range $-35$ to $-30$, to accommodate the higher airmass and maintain a good SNR. With a seeing between 0.55\arcsec and 0.65\arcsec, this ensures a SNR of 3 for all our $z_{ab}^\prime$-band observations with a limiting magnitude of $z_{ab}^\prime$=24. 

The WIRCam images were reduced using the method described in \cite{albert_37_2011}. First, they were preprocessed by CFHT using their `I`iwi pipeline version 2.0.
Next, a low-pass filter was created by median binning the image by 4x4 pixels, applying a 5x5-pixel median filter, and then re-sampling at the original image size. This low-pass filter was subtracted from the image to preserve only high spatial frequencies. After this, the different images were stacked using the Bertin software suite. First, SExtractor \citep{bertin_sextractor:_1996} builds a catalog of objects in each image. This catalog is in turn read by Scamp \citep{bertin_scamp:_2010}, which also computes the astrometric and photometric solutions by anchoring on the $J$-band data of the 2MASS catalog. Swarp \citep{bertin_swarp:_2010} then stacks the images together. 

Data from MegaCam were first processed by CFHT's Elixir pipeline. Then, the astrometric solution from each of the 40 CCDs was found by anchoring the field on the positions from the USNO-B1 catalog. A high-pass filter was applied, as before, by subtracting an image created by median binning the image by 4x4 pixels, applying a 7x7-pixel median filter, and then re-sampling at the original image size. Then the images from the different CCDs were combined to form an image of size matching that of the WIRCam images, as the field of view of MegaCam is much wider than WIRCam's. The different images of a given target obtained on a given night were then median-combined to obtain the final reduced $z_{ab}^\prime$ image.

\subsubsection{$Spitzer$/IRAC observations}

Our complete sample was observed with $Spitzer$/IRAC. Nine of our targets had previously been observed with IRAC with an exposition time that meets our requirements. The others targets were observed between 2015 and 2016 ($Spitzer$ proposal 11092) in both IRAC $[3.6]$ and $[4.5]$, with a total integration time of 2160~s in each band (per-visit total of 5221~s with overheads). More precisely, we used 30~s individual exposure time, two exposures per dither position per band, and a 36-exposure reuleaux dither pattern.

The Spitzer/IRAC pipeline reduced images were further processed with custom IDL routines. First, the different images of a given target were oversampled on a $0.5\arcsec$ pixel grid and median-combined using the pipeline-provided astrometry and polynomial distortion. Then, to preserve the PSF morphology orientation in the image, in view of the PSF subtraction routines to be applied, we registered all images to a common PSF rotation angle. 

Further data reduction involved the subtraction of the stellar point spread function (PSF) to reveal embedded and close-in sources. Since the \textit{Spitzer} observations were uniform, we used the Reference Differential Imaging technique to subtract the PSF from a reference library. The strategy is similar to the re-analysis of \textit{Hubble} imaging data through the ALICE project \citep{soummer_archival_2016,choquet_archival_2015}, and to previous analysis of archival \textit{Spitzer} data \citep{janson_high-contrast_2015,durkan_high_2016}.

The library of reference PSF was created out of the newly obtained data, using the PSF of the observed stars. Saturated stars, very crowded fields, and low-contrast ($<1$) visual binaries were removed from the library, resulting in a total of 111 PSFs out of the 168 targets observed. Each image was registered on a common center based on the fit of a two-dimensional Moffat function. It was then normalized in brightness from the flux measured in an aperture of a radius of seven pixels centered on the PSF core. Point sources were identified as $3\sigma$ outliers from the noise (calculated with a robust sigma estimator) after an initial PSF subtraction from the median of the reference library, excluding the given image. They were subsequently masked out in the original image.

We used classical RDI to subtract stellar PSFs in each \textit{Spizter} image for both filters. Because of the very large number of point sources in our deep data, advanced techniques such as LOCI or PCA \citep{lafreniere_new_2007,soummer_detection_2012} suffered from too many pixels that were masked out, reducing the effective number of reference images. They tended to oversubtract the target PSF and other point sources in the field. We therefore opted for a classical median subtraction, a trade-off between the quality of the PSF subtraction and source preservation. Following this strategy, the processed image under consideration was excluded from the library of references, the median of which was then taken as the reference PSF for subtraction of this image. The position of the star was estimated by fitting a Moffat function and the reference PSF was shifted to this position. The reference PSF was normalized to the target brightness within the same aperture and subtracted from the image. This three-step process was repeated for any low-contrast ($<1$) visual companion of our target present in the field. For tight binaries or triple systems, the subtraction of all PSFs was done at once by iterating over the position and flux of each component in order to minimize the residuals in a box of width of 30 pixels. Saturated stars were processed like binaries to optimize the star registration and flux normalization. They still suffered from poorly subtracted wings and bright vertical stripes escaping from the core. Therefore, a new library was built from residual images of similarly saturated stars, following the same cleaning processes as for the original library. The median of this new library was used to subtract these residuals. 

The images were finally high-pass filtered by subtracting a median filter of width 15 pixels.

\subsubsection{Archival $Spitzer$/MIPS 24~$\mu$m data}

A search through the $Spitzer$ archive revealed that 141 of our 177 targets were observed with MIPS at 24 $\mu m$ as part of surveys to find infrared excess indicative of debris disk. A flux measurement (or upper limit) at such a longer wavelength can be useful to better constrain the SED of our candidate objects identified. Thus for those 141 targets we retrieved the MIPS data and built our own combined image using the Enhanced BCD images (EBCD), as they have a superior flat fielding than the BCD image. A high-pass filter was then applied by subtracting a double-pass median-filtered version of the image, respectively of widths 32 and 12. We obtained the limiting flux in Jy/sr by doing aperture photometry at several random positions over the whole image and then evaluating the robust standard deviation of the resulting flux distribution. We converted this limiting flux in Jy/sr to magnitude and, over all images, obtained a median limiting magnitude of 12.5~mag.


\subsection{Photometric calibration}

Our GEMINI observations were all acquired with a specification for observing conditions of up to 70\% cloud cover, or patchy clouds. Under those conditions, a variation of up to 0.3 mag can be expected. We assessed if significant variations were present or not from the data themselves. For a sequence of observations of a given field, we calculated the standard deviation of the flux variations of the 20 brightest stars, as compared to a reference image from the sequence. If this variation was higher than 3\%, then we considered that the images of that target were not taken in photometric conditions ('patchy clouds' in table~\ref{zpointJ} and table~\ref{zpointz}); otherwise we considered that the images were taken under photometric conditions ('phot' in table~\ref{zpointJ} and table~\ref{zpointz}).

All of our CFHT images were taken in good conditions, with seeing around 0.6 for MegaCam and 1.1 for WIRCam, but we still checked the flux variations between images for a given target on a given night to make sure that the observations were acquired in photometric conditions.  

Our final, stacked images in $z_{ab}^\prime$ and $J$ were calibrated in flux by comparison with the Sloan Digital Sky Survey catalog \citep[SDSS DR9;][]{ahn_ninth_2012} or the 2MASS All-Sky Catalog of Point Sources \citep{cutri_2mass_2003}, respectively. If SDSS data were not available for a given field, we used either PanSTARRS \citep{chambers_pan-starrs1_2018} $z_{p1}$ data (available for 55 of our targets with dec $>$$-30$) or SkyMapper  \citep{wolf_skymapper_2018} $z^\prime$ data (available for 79 of our targets). The PanSTARRS filters ($g_{p1}$, $r_{p1}$, $i_{p1}$, $z_{p1}$, $y_{p1}$, $w_{p1}$) are not the same as the SDSS filters, so we used the \cite{tonry_pan-starrs1_2012} color corrections to convert the Pan-STARSS magnitudes to SDSS magnitudes. For the $J$ band, if too few stars in our images were in the 2MASS PSC, we used deeper data from the VISTA Hemisphere Survey \citep[VHS;][]{mcmahon_first_2013} or archival observations from the Observatoire du Mont Mégantic obtained using the Spectrographe infrarouge de Montréal (SIMON) \citep{albert_recherche_2006}.

For a given image, the magnitude that produces one count per second on the detector, or the zero point, was calculated for each individual point source in common between our image and the catalog, based on the difference between the magnitude extracted from our image and the magnitude taken from the catalog. Then, the zero point of the image was taken to be the median of the individual zero points, and the error was computed by taking the standard deviation of those individual zero points divided by the square root of the number of sources. 

When no catalog data were available for a given image, or when less than 5 objects with a magnitude measurement were available in the field of view, we used as a zero point the median zero point for the given observing condition ('phot' or 'patchy clouds'), instrument and filter of the image. This occurred for 16 of our $J$-band images and 48 of our $z_{ab}^\prime$-band images. In the $J$ band, we obtained a zero point of 22.4 $\pm$ 0.7 and 22.1 $\pm$ 0.9 with Gemini/F2, and 22.5 $\pm$ 0.8 and 22.7 $\pm$ 0.6 with CFHT/WIRCam, for photometric and non-photometric conditions respectively. In the $z$-band, we calculated a median zero point of 24.5 $\pm$ 0.3 and 24.6 $\pm$ 0.5 with CFHT/MegaCam and 29.7 $\pm$ 2 and 28.9 $\pm$ 2.3 with Gemini/GMOSS, for photometric and non-photometric conditions respectively.

\subsection{Follow-up observations}
Our search for planetary mass object revealed a number of candidates (see Section~\ref{follow}) that motivated us to obtain follow-up observations. 

An astrometric follow-up was carried out between 2016 and 2017 in the $J$-band, with either CFHT/WIRCam or Gemini-South/Flamingos2. Only $J$-band images were obtained as it is in this band that the SNR of the candidate is highest. We used the same observation parameters as for the first epoch observations. We obtained proper motions follow-up in the $J$ band for 4 candidate companions. 

%


\section{Results}\label{results}  
The ground-based observations described earlier were designed to reach a limiting magnitude of $z=24$~mag at $3\sigma$ and $J$=21~mag at $7\sigma$. In practice, we achieved a median [AB] limiting magnitude in the $z$-band of $23.4\pm1.2$~mag with CFHT/MegaCam and $23.7\pm1.2$~mag with Gemini/GMOS-S, at $3\sigma$. In the $J$-band, we achived a $7\sigma$ median  Vega limiting magnitude of $21.2\pm0.5$~mag with CFHT/WIRCam and $21.0\pm0.8$~mag with Gemini/F2. For the $Spitzer$/IRAC observations, we reached a median magnitude limit of $18.5 \pm 0.9$~mag at [3.6] at $5\sigma$ and $18.5 \pm 0.8$~mag at [4.5] at $3\sigma$.

\subsection{Detection Limits}

\begin{figure*} 
\centering
\plottwo{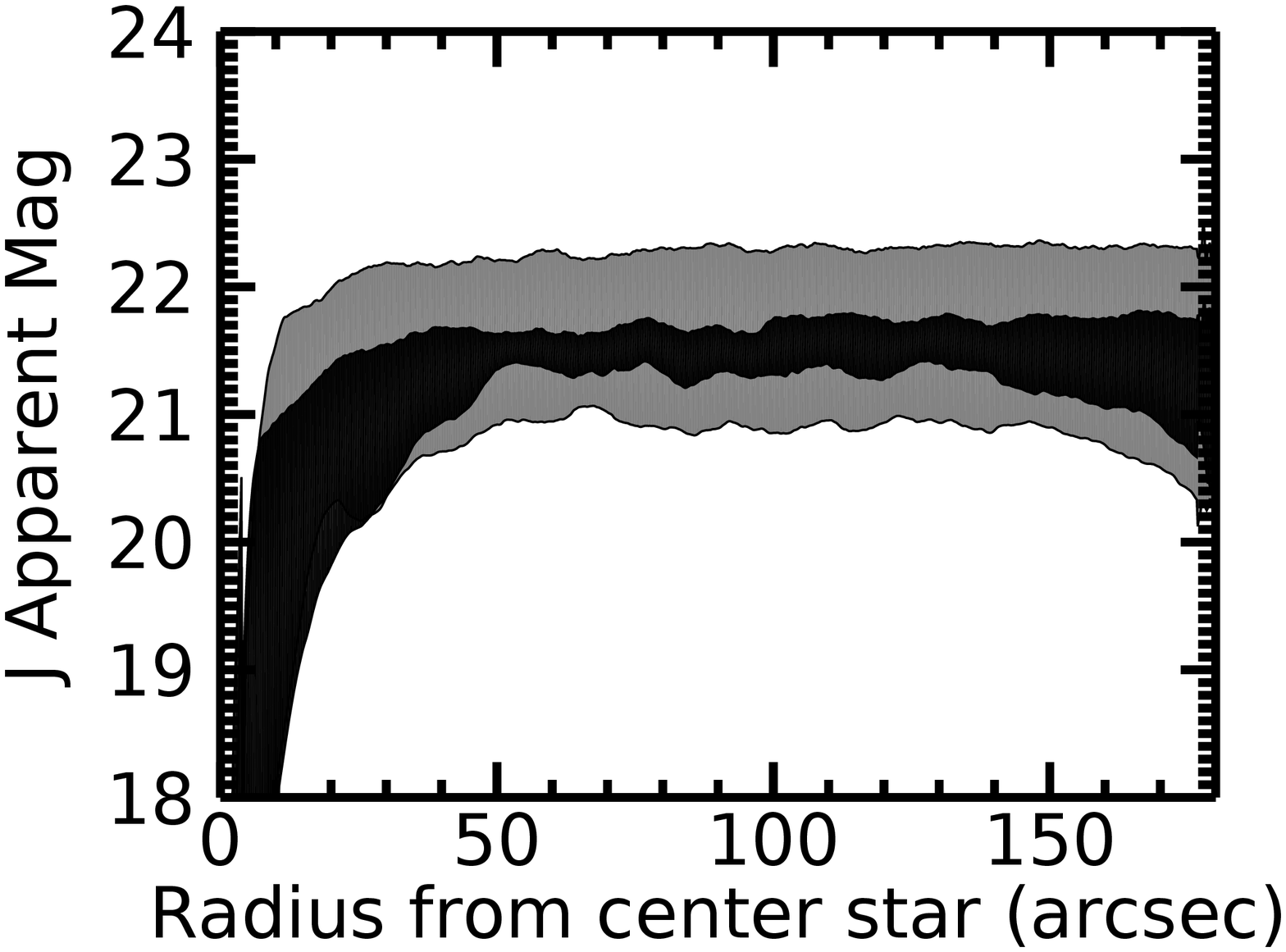}{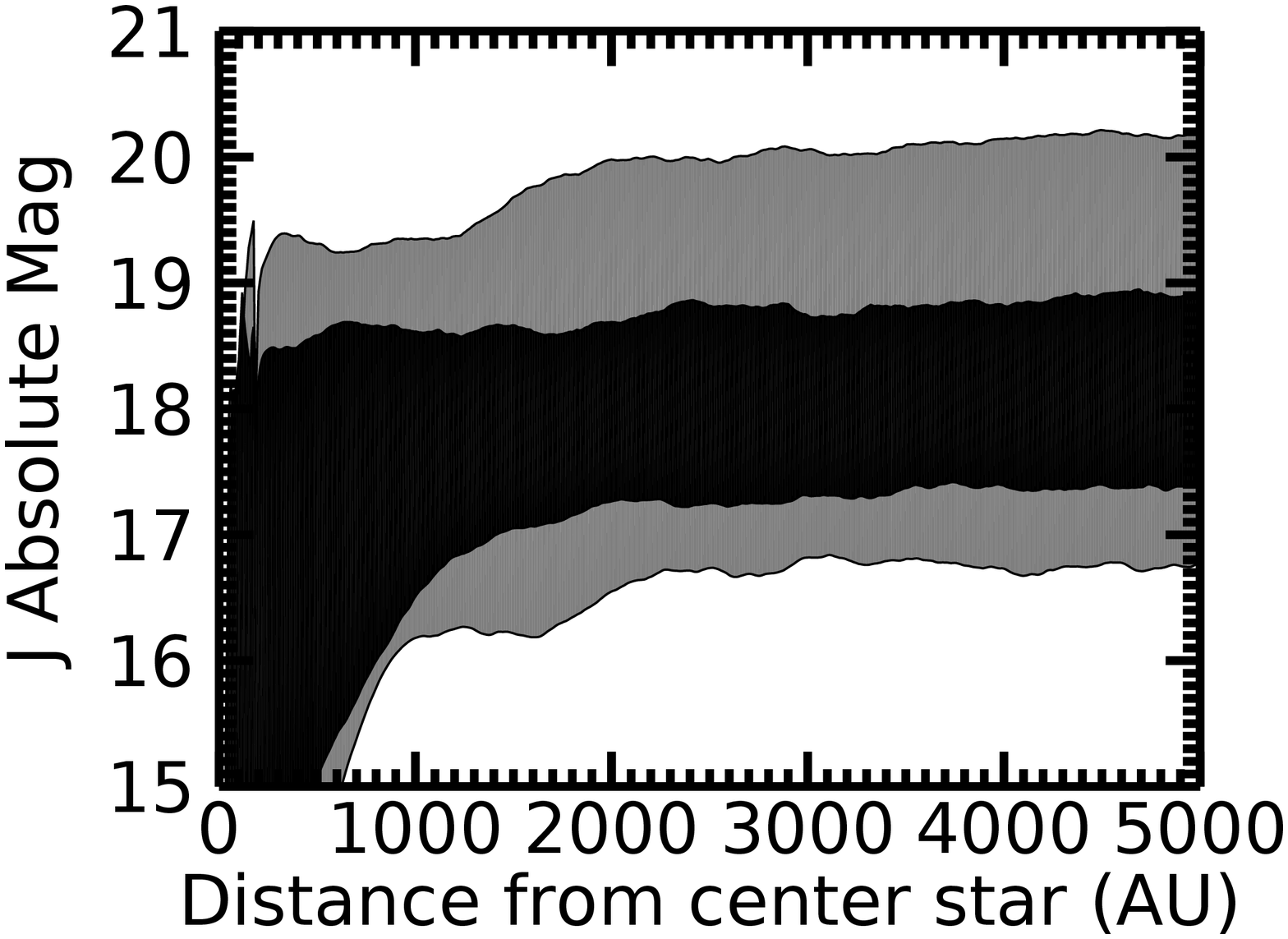}
\caption{\label{declimF2} Detection limits for all of our stacked $J$-band images observed with Flamingos-2 at Gemini-South or WIRCam at CFHT. The left panel shows limiting apparent magnitudes as a function of the projected separation from the target star in arcseconds. The right panel shows the corresponding absolute magnitudes at the distance of the star as a function of projected separation from the star in AU (with a cutoff at 5000 AU). 50\% of the detection limit curves fall inside the black region while the grey region contains 80\% of the curves. }
 \end{figure*}

\begin{figure*} 
\centering
\plottwo{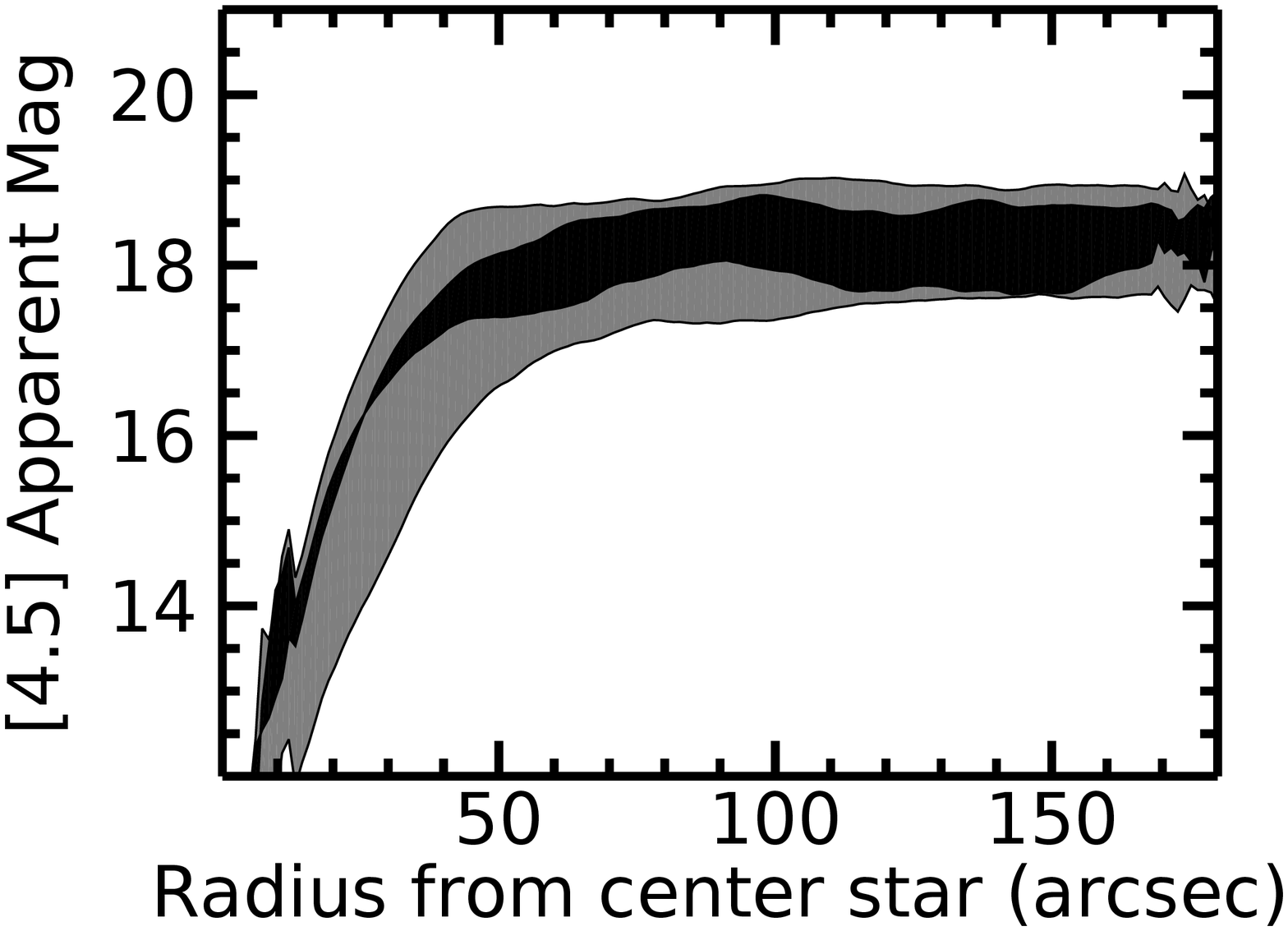}{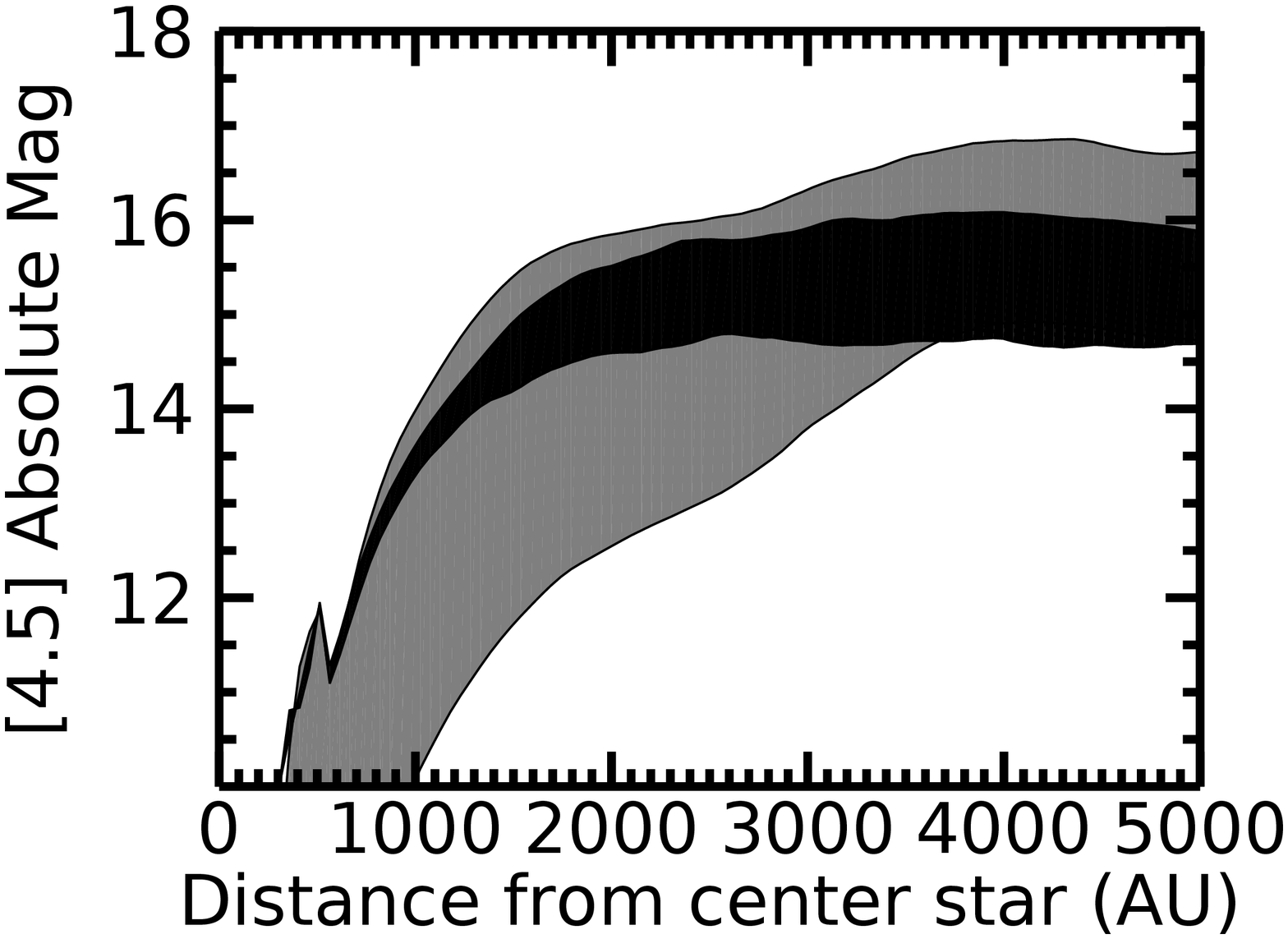}
\caption{\label{declimspitzer} Same as~\ref{declimF2} for the $Spitzer$/IRAC observations. }
 \end{figure*}

The sensitivities to companions, in terms of limiting magnitudes, were evaluated for each $J$-band stacked image and [4.5] image as a function of the radial distance from the target star. For each radius from the central star, aperture photometry was performed by obtaining the flux inside 100 apertures of radius of 1 FWHM and a sky annulus between 4 to 6 FWHM.  The limiting flux at each radius is the standard deviation of these 100 fluxes and it was then converted into magnitudes to get these $7\sigma$ limiting magnitudes. Theses results are presented in Figure~\ref{declimF2} for the $J$-band images and in Figure~\ref{declimspitzer} for the [4.5] images. They show that the limiting magnitudes initially grow with increasing distance from the central star, and then reach a plateau. The black region of the plots contain 50\% of the detection limit curves while 80\% of the curves fall inside the grey region. For the $J$-band images, the plateau is reached at $\sim$30\arcsec\ at a magnitude of $J\sim 21.5$~mag for 50\% of our target stars (the black region). The curves are truncated at about 180\arcsec, which corresponds to the limit of the field of view of the Flamingos-2 images.  The plateau is reached at a projected physical separation of 1000~AU for an average star of our sample. For our [4.5]  images, the plateau at magnitude $\sim$18.5 is reached at a radius of $\sim$50\arcsec for 50\% of the stars of the sample (the black region). We used the same cut-off as the $J$-band images. Tables~\ref{liste_detect_lim} and \ref{liste_detect_lim_spit} present, respectively for the $J$-band and [4.5] images, the $7\sigma$ detection limits for each target over the plateau along with the minimum and maximum separations (in arcsec and AU) where these limits are valid (defined as the range for which the detection limit is at most 1 magnitude brighter than the plateau value given, to accommodate for small fluctuations with separations).

The limiting magnitudes can be converted to limiting masses using evolutionary models at the ages of our targets (which range between 10 and 150~Myr). We used the COND models from \citet{baraffe_evolutionary_2003} to infer the masses. These models assume a hot start, which as described by \cite{bowler_imaging_2016}, corresponds to idealized initial conditions and an arbitrarily large initial radius. This model is thus optimistic as it represents more luminous planets than cold start models. The mass limit reached over the sensitivity plateau for each target is indicated in Table~\ref{liste_detect_lim}  and \ref{liste_detect_lim_spit}.

\subsection{Candidate Search}

\begin{figure*}[ht]
\centering
\plotone{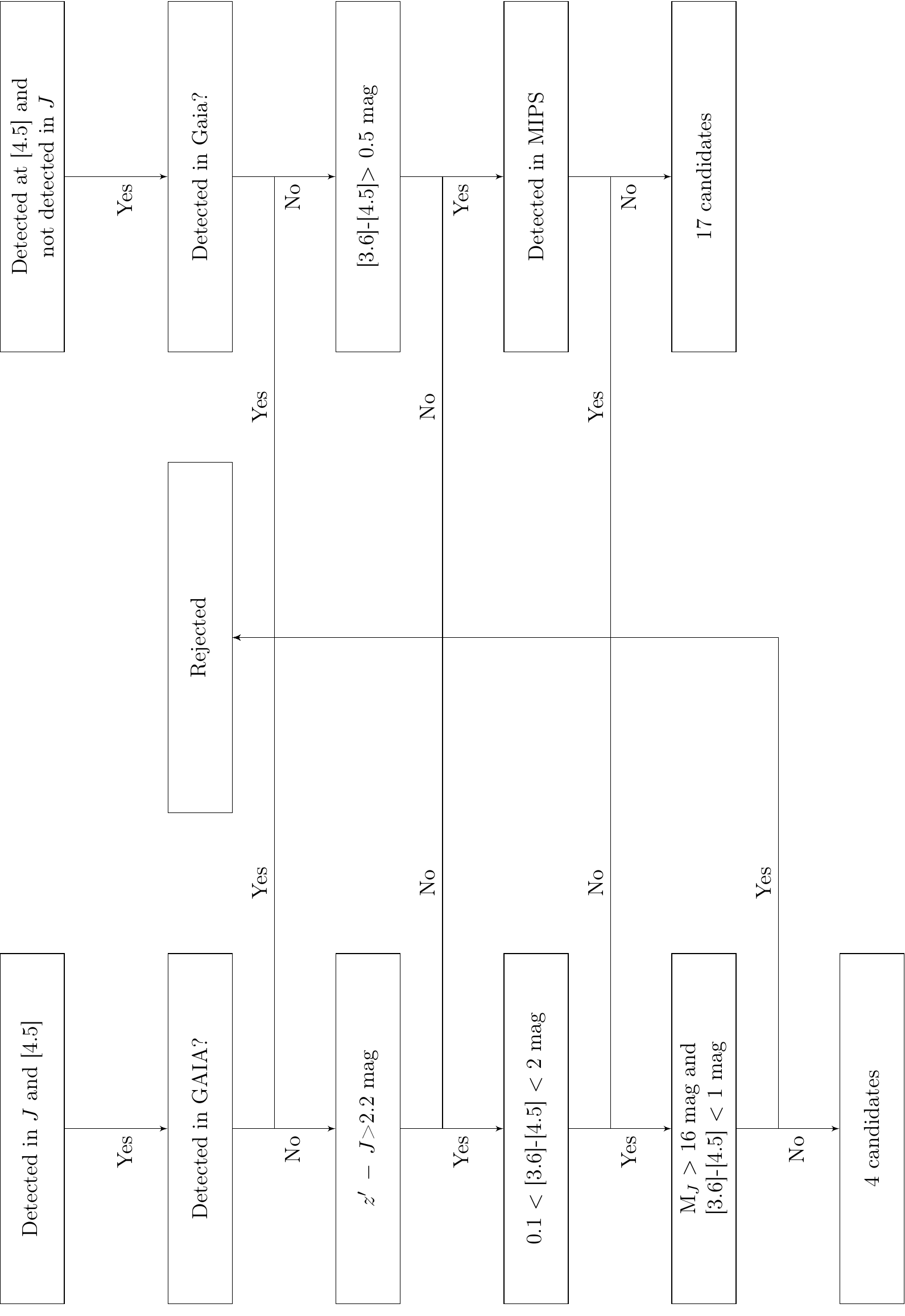}
\caption{\label{organizJ}
Flowchart of the candidate selection for candidates detected in the $J$-band on the left and the IRAC-only candidates on the right.
}
\end{figure*}

We searched for and identified candidates in our imaging based on their $z_{ab}^\prime-J$ and $[3.6]-[4.5]$ colors. We started by identifying all point sources in the $J$-band images using the IDL {\tt find} procedure (from Astrolib) and then fitted a 2D Gaussian function to each of them to get a more precise position. At this step, sources with an elongated PSF were rejected, as a first attempt to exclude extra galactic contaminants. We also rejected sources too close to the edge of the field (for F2 or GMOS-S) and sources that were saturated in either band. We used coordinates measured in our $J$-band images to identify sources in the $z_{ab}^\prime$-band images. In both bands, we used aperture photometry with a radius of 1~FWHM and a sky sampling annulus extending between 2 and 3~FHWM to retrieve the instrumental flux of each source.
 We kept only point sources detected at $7\sigma$ in $J$, $5\sigma$ in [4.5] and $3\sigma$ in [3.6].
At a distance of $>20$~pc, which is the case for 90\%  of the stars in our sample, a radius of 5000~AU fits in the field of view of the $Spitzer$/IRAC images. For that reason, we searched for candidates only inside a projected separation of 5000 AU from the target stars.

We found the center of the target star by fitting a 2D Gaussian to the PSF, for stars that were not saturated. However, most of our targets were saturated in our $J$-band images. Thus, we used the  \emph{Gaia}~DR1 DR1 catalog \citep{gaia_collaboration_gaia_2016} to find an approximated position of the star. We then used the known proper motion of the star to compute the position at the time the image was obtained. If  \emph{Gaia} data were not available, we fitted a 2D Gaussian to the PSF, where all of the saturated pixels were given the maximum value possible for a pixel. For the IRAC images, the center of the stars was obtained during the PSF removal process.
 
\subsubsection{Colors}
For the ages of our target stars, 10--150~Myr, the transition between brown dwarfs and planets happens between L1/L2 and L5/L6 based on AMES.Cond models \citep{baraffe_evolutionary_2003}. As mentioned above (and see Figure~\ref{col_spt}), early-type L dwarfs have a $z_{ab}^\prime-J$ color $\gtrsim$2.5~mag. Considering our errors on magnitudes and zero points, we selected only sources with $z_{ab}^\prime-J>2.2$. Per the above discussion, this same color cut is also sensitive to T and Y dwarfs,
which can be identified either through detection in both bands or as $z_{ab}^\prime$ dropouts (in the cases without detection in $z_{ab}^\prime$, we get only a lower limit on the $z_{ab}^\prime-J$ color).Thus at this stage, we kept all sources with $z_{ab}^\prime-J > 2.2$~mag, including all the $z_{ab}^\prime$ dropouts.

As a second step, we removed any source that has a counterpart in the  \emph{Gaia}~DR1 DR1 catalog \citep{gaia_collaboration_gaia_2016}.  \emph{Gaia} can detect objects with magnitude as low as $G$=20. Based on the relationship between $G-J$ and the spectral type of L dwarfs from \cite{smart_gaia_2017}, using the expected $J$ magnitude of L dwarfs from \cite{faherty_population_2016}, and assuming a distance of 42 pc (the median distance of our sample), the cut in  \emph{Gaia} magnitudes rejects objects earlier than $\sim$L2.  

\begin{figure}[t]
\centering
\includegraphics[width=\linewidth]{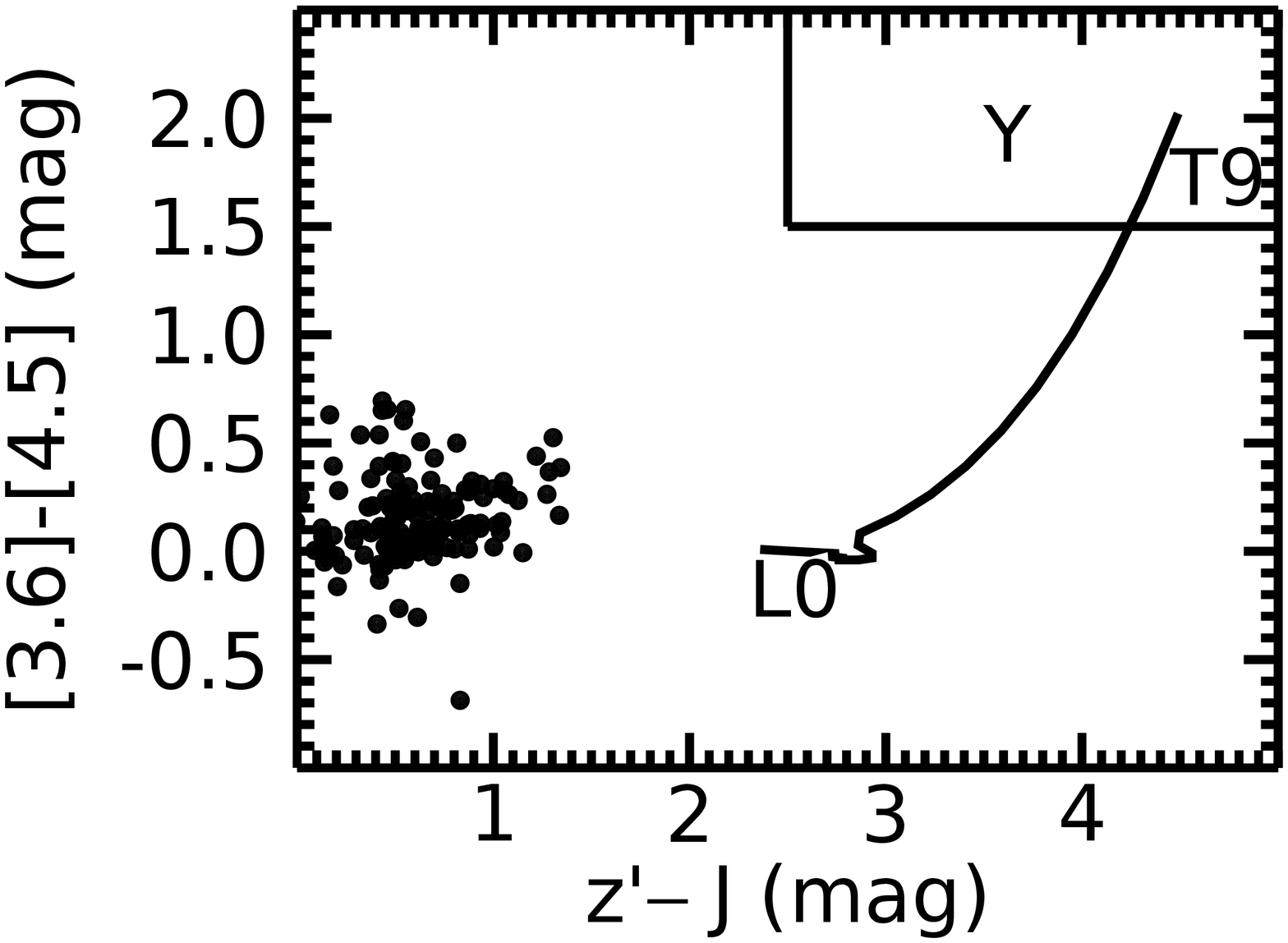}

\caption{\label{colortarget179} Color-color diagram for HIP 26453, a known member of Columba. The dots represent all sources detected in our $J$-band imaging, and without detection in  \emph{Gaia}, within a radius of 5000 AU from the target star. The solid line shows the expected color sequence for spectral types L to T from \cite{dupuy_hawaii_2012}. The box represents the expected colors for early Y dwarfs. No candidates were detected in this field.   }
 \end{figure}

Then, we compared the $z_{ab}^\prime-J$ colors and $[3.6]-[4.5]$ colors of our candidates to typical colors of ultracool field dwarfs \citep{dupuy_hawaii_2012}, see Figure~\ref{colortarget179}. This figure shows all point sources in a radius of 5000 AU in the $J$-band image for an average target of the sample for which there was no candidate detected. The solid black line represents the expected colors for L to T dwarfs according to \cite{dupuy_hawaii_2012}. We kept as candidates only the detections with $[3.6]-[4.5] \sim 0.1$ to 2 ~mag, as this is the expected interval for T dwarf's colors. We also kept as candidate source with M$_{J}$ $<$ 16~mag and $[3.6]-[4.5]$ < 1~mag or M$_{J}$ $>$ 16~mag and $[3.6]-[4.5]$ > 1~mag. Figure~\ref{organizJ}, right, presents the flowchart of the candidate selection for the candidates detected in the $J$ band.

\begin{figure}
\centering
\epsscale{1.1}
\includegraphics[width=\linewidth]{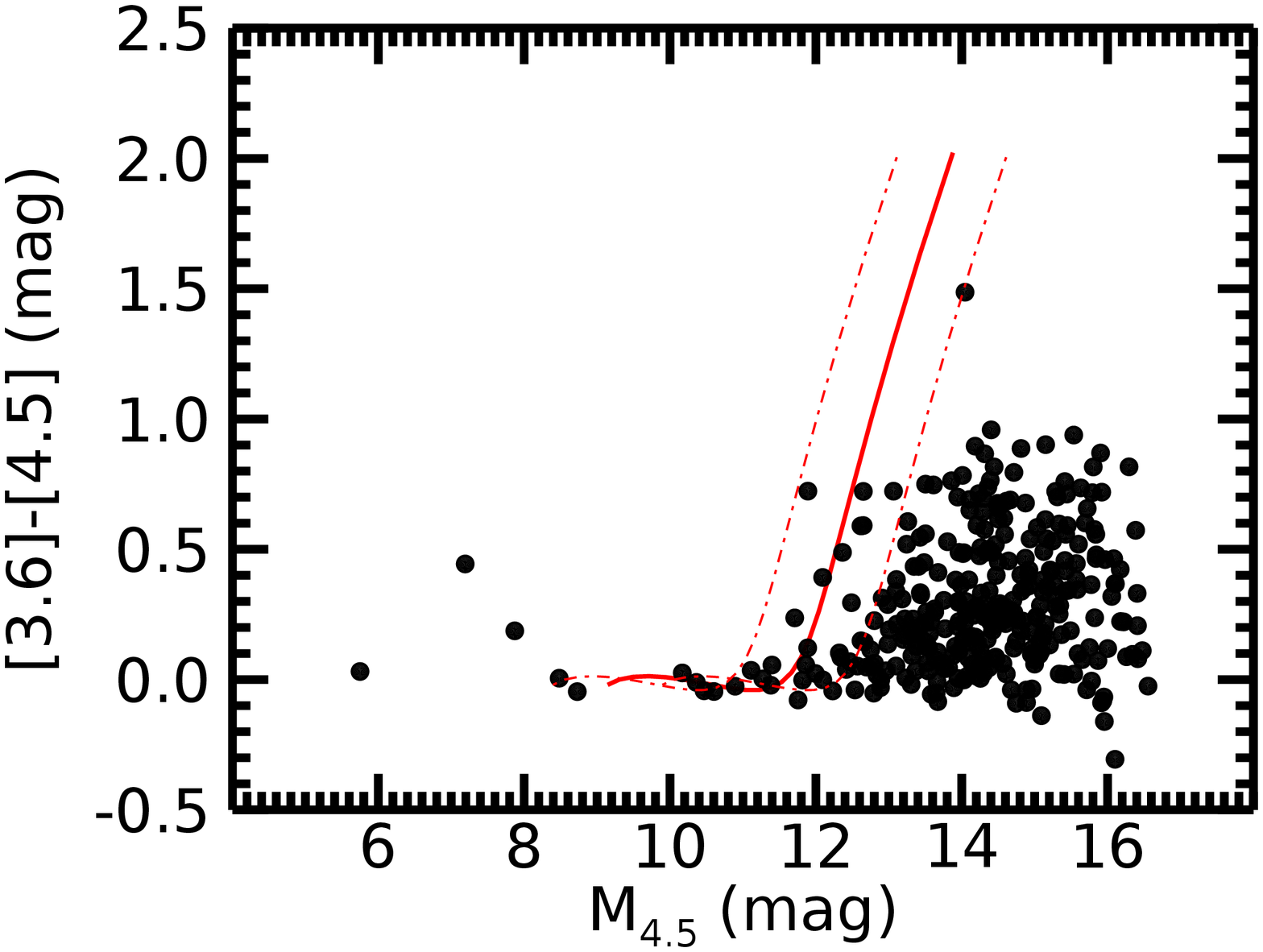}
\caption{\label{candspitzer} $[3.6]-[4.5]$ color of sources detected in our Spizter imaging of HIP 11152 versus their $[4.5]$ absolute magnitude at the distance of the target star. The solid red line corresponds to the colors of M6 to T9 dwarf from \cite{dupuy_hawaii_2012}. The dotted lines on either sides represented a spread of 0.75 magnitude. The dots are all the point sources presents in a sphere of 5000 AU around the central star for which there is no detection in the optical. One point source has colors consistent with a late T dwarf at the right absolute magnitude. This point source is not detected in the $z_{ab}^\prime$ nor $J$ images. While it is expected for a planetary mass companion to be undetected in $z_{ab}^\prime$, it should have been detected in $J$ images, given our detection limits. It is thus likely that the candidate is in fact an extragalactic contaminant.   }
 \end{figure}

In some cases, a source was detected at $5\sigma$ in our IRAC data but we found no counterpart in our $J$ or $z_{ab}^\prime$ imaging, respectively at 7 and $3\sigma$. Unambiguous IRAC-only detection of planets is possible only if $[3.6]-[4.5] > 2$, which corresponds to our  detection limits of $\sim$21 in the $J$ band, or M$_J\sim 18$ (T8.5) at 50~pc according to AMES.Cond models. However, most of our IRAC-only detection had $0.5<[3.6]-[4.5]<2$. As the color in those bands for young 2 Mjup objects is rather uncertain, we decided to follow-up these sources anyway. Thus from the IRAC-only detection, we selected only sources with $[3.6]-[4.5]>0.5$ and no \emph{Gaia} detection. In addition, as the absolute magnitudes of young planetary mass objects analog to T dwarfs are not well known, we kept only sources with a $[4.5]$ absolute magnitude within 0.75~mag from the typical values of field T dwarfs, see Figure~\ref{candspitzer}. This method has uncovered 79 candidates with the expected colors of T dwarfs. Figure~\ref{organizJ}, on the right, presents the flowchart of the candidate selection for the candidates not detected in the $J$ band.

The color criteria above yielded typically a few candidates per field. However, most were easily discarded by either looking at the stacked images or the individual frames: some had an elongated PSF that escaped our automatic cut, some fell out of the detector in one or more frames of the dither pattern biasing their photometry,  some were due to a persistence signal from a bright star that was on the same part of the detector in a previous frame (for the WIRCam images), and some fell over the spider diffraction spikes of the host star. After these initial verifications, our search yielded 4 candidates with $J$-band detection and 48 candidates with IRAC-only detections.

\subsubsection{Cross-match with the 2MASS calatog}
The detection method described earlier is not sensitive to companions with spectral type earlier than early L dwarf. Instead, the latest M to early L-type dwarf companions can be identified through a search for common proper motion based on a comparison of our $J$-band images with 2MASS images, given the $\sim$15 years baseline between them. We performed such a proper motion comparison for all sources with $J$<16.5~mag. 

This search identified one candidate with a proper motion consistent with a target star. It is TWA30B, an M4V dwarf companion of TWA30 -- an M5 dwarf member of the TW Hydrae association -- at a separation of 3400 AU and which was discovered previously by \cite{looper_widely_2010}.

\subsubsection{Follow-up of candidates}\label{follow}
\begin{figure}[h] 
\centering
\includegraphics[width=\linewidth]{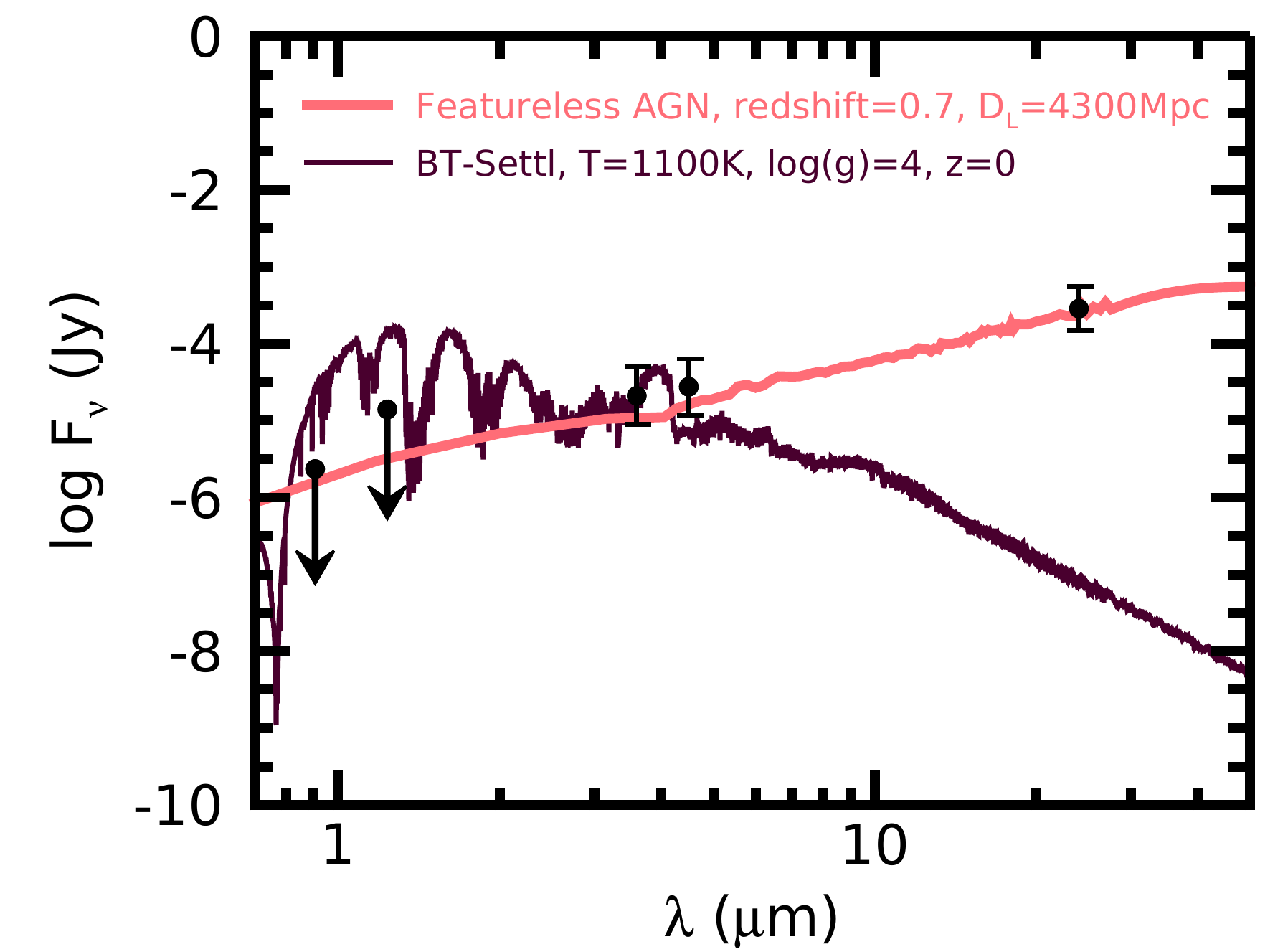}
\caption{\label{agn} Photometric data for one candidate that has a large $[3.6]-[4.5]$ color but no detection in $z_{ab}^\prime$ and $J$. The data are compared to the model spectrum of an object with a $T_{\rm eff}=1100$~K, $\log{g}=4$ and $z=0$ from BT-Settl (purple) and to the spectrum of a featureless AGN with a redshift of 0.7 and a D$_L$=4300~Mpc \citep[magenta, from ][]{kirkpatrick_goods-herschel:_2012}. We see that the detection at 24 $\mu$m makes it very easy to untangle between a mid-T dwarf and a AGN. 
}
 \end{figure}   
\begin{figure*}[ht] 
\includegraphics[width=\textwidth]{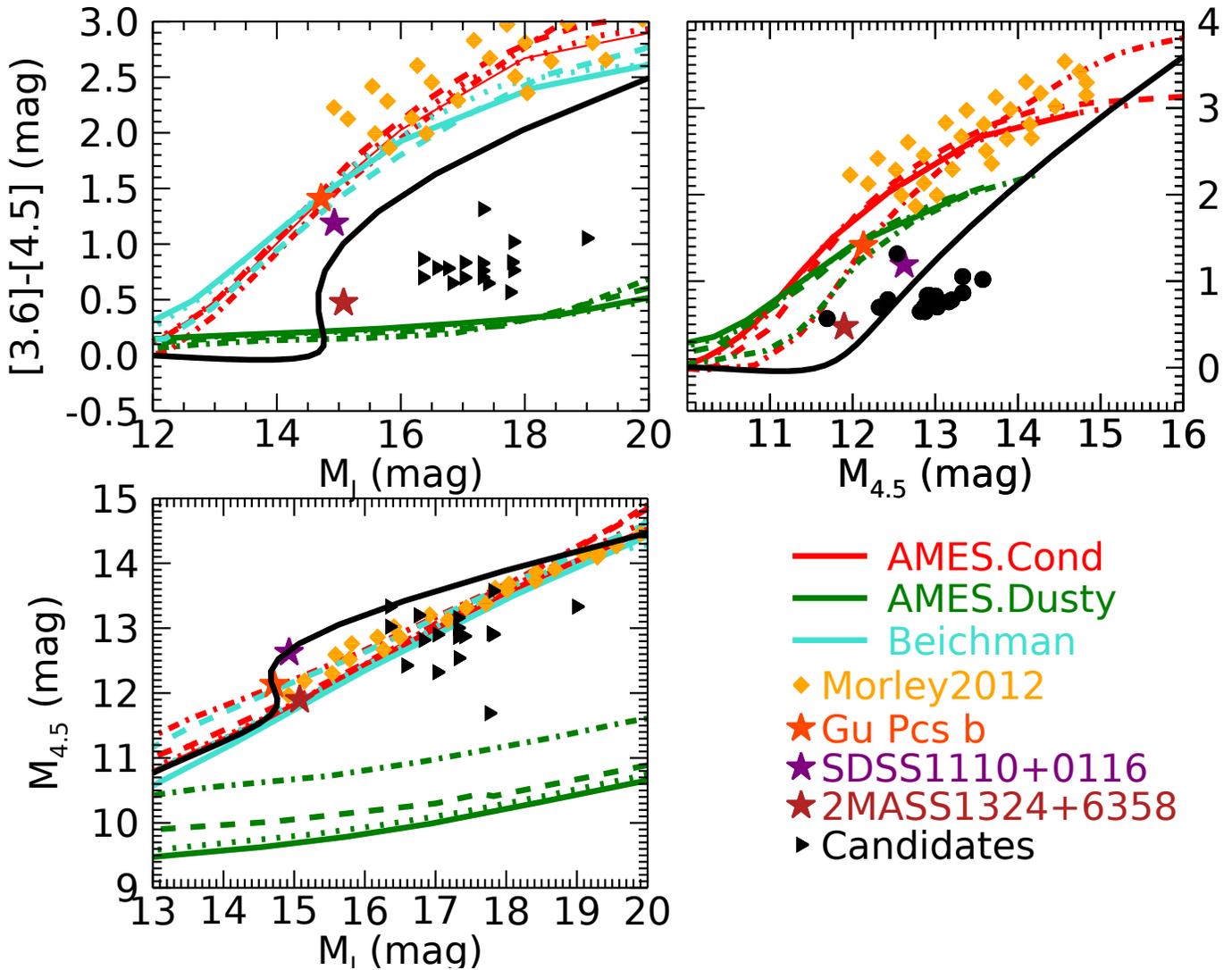}
\caption{\label{candspitzer_all} Colors of our 17 $Spitzer/IRAC$-only candidates remaining after the MIPS detection cut (triangles, upper limits in $J$-band). $[3.6]-[4.5]$ colors versus absolute $J$ magnitude are shown on the upper left while $[3.6]-[4.5]$ colors versus absolute [4.5] magnitudes are displayed on the upper right. The the lower left shows absolute [4.5] magnitudes vs absolute $J$ magnitudes. Colors for M6 to T9 dwarfs from \cite{dupuy_hawaii_2012} are shown with a black line. The red curves represent the Ames.Cond models \citep{baraffe_evolutionary_2003} at 10, 20, 120 and 5000~Myr, using respectively the solid, dotted, dashed and dash-dotted line. Also shown are models from \cite{beichman_wise_2014} in cyan, \cite{mordasini_characterization_2012} in yellow and Ames.Dusty in green. Photometric data for 3 young T dwarfs are also shown by an orange star for Gu Psc b \citep{naud_discovery_2014}, a purple star for SDSS1110+0116 \citep{gagne_sdss_2015} and a red orange star for 2MASS1324+6358 \citep{gagne_2mass_2018}. While the candidate companions have similar [3.6]-[4.5] colors versus [4.5] as the young T dwarfs, they are too faint in the $J$-band to be considered planetary objects. 
}
 \end{figure*}

The follow-up of our candidate companions includes 3 different types of observations. First of all, IRAC-only detections were studied in greater details by using MIPS data. A photometric follow-up was obtained to try to identify puzzling objects with very red $[3.6]-[4.5]$ colors and no detection the $z_{ab}^\prime$ and $J$ bands. Lastly, a proper motion follow-up was obtained for all candidates detected in $J$ that survived the color cuts and verifications.

Our search for candidates in the $Spitzer$/IRAC images yielded 48 candidates with $[3.6]-[4.5]> 0.5$ and no detection in $z^\prime$ or $J$.  Figure~\ref{candspitzer} shows all the point sources detected in $[3.6]$ and $[4.5]$ in a given field and for which no visible counterpart was found (from the  \emph{Gaia}~DR1 catalog). Faint, red objects like this certainly constitute interesting planetary mass candidates, as indeed it is expected for such objects to be $z^\prime$ dropouts. Yet given our limits it is unexpected for them to be unseen in $J$. Other astrophysical sources that may have similar photometric properties include galaxies and active galactic nuclei (AGNs).

Figure~\ref{agn} shows the expected SED of a low-mass object with an effective temperature of 1100~K compared to the SED of a featureless AGN. As the Figure illustrates, it is difficult to untangle AGNs from planetary candidates using $[3.6]$ and $[4.5]$ photometry alone, but photometry at 24~$\mu m$ is a very good discriminator. 
We used the MIPS 24 $\mu m$ images mentioned above, reaching a limiting magnitude of 12.5 at $1\sigma$ in that band for most targets, to see if our candidates were detected at that wavelength, which would be incompatible with a planetary mass object. This enabled us to reject 31 of our remaining IRAC-only candidates and to identify them as extra-galactic contaminants. We checked archives to see if those MIPS detection are associated with X-ray or radio emission, but none of them are already known as AGN.

After this cut, 17 IRAC-only candidates remain. Figure~\ref{candspitzer_all} shows the colors and magnitudes of the candidates compared to different models as well as to photometric data from known young T dwarfs. Of those 17 candidates, 4 were observed by MIPS but not detected. These candidates have $[3.6]-[4.5]$ = 0.7 to 0.9~mag and $[4.5]$ magnitudes between between 15.8~mag and 17.5~mag. Using only their IRAC color and assuming that they are T dwarfs and that the BT-Settl/Ames.Cond model are valid, one would expect them to have M$_J$ $\sim$ 15~mag, which would have been detected by our survey. As these candidates show no detection in our $J$-band imaging, we rejected those 4 candidates. The last 13 candidates were not observed by MIPS. Those candidates have $[3.6]-[4.5]$ colors between 0.6~mag and 1.3~mag and [4.5] magnitudes between 15.2 and 16.9~mag. Using the same thought process as for the candidates not seen in MIPS, we see that those candidates also should have been seen in the $J$ band, but they were not detected. We thus reject the last 13 candidates such that no IRAC-only candidate remain. However, we decided to list those 17 rejected candidates in the interest of completeness, as we cannot identify the nature of the candidates at this stage, and because models might not reproduce accurately the colors of young late T to early Y dwarfs. Table~\ref{liste_extra} lists them all, with their RA, DEC, associated host star, limiting magnitude in $z^\prime$ and $J$, apparent magnitude in $[3.6]$ and $[4.5]$, separation in AU from the host star, and distance of the host star in pc. These unknown objects are possibly Ultra Luminous Galaxies (ULIRGS). ULIRGS are identified by their red [3.6]-[4.5] $>$ 0.5 colors meaning that they share colors with T dwarfs. \cite{daddi_multiwavelength_2007} have shown that ULIRGS from the GOODS sample, with 0.7$<$z$<$1.3 have a space density of $2x10^{-5} Mpc^{-3}$. At a luminosity distance corresponding to a redshift of z=1, about 3 ULIRGS should have been found per $Spitzer$/IRAC field. As ULIRGS have F$_\nu  \sim 10\mu Jy$ for $z \sim $ 1 to 2 \citep{kirkpatrick_further_2012}, they are expected to be detected in our images.

A proper motion follow-up was obtained for all 4 candidates identified through their $z^\prime - J$ and $[3.6]-[4.5]$ colors. It was carried out between 2016 and 2017 both at CFHT and at Gemini-South. Table~\ref{listecand} lists the candidates with their RA, DEC, host stars, M$_{z^\prime}$, M$_{J}$, M$_{3.6}$, M$_{4.5}$, separation in AU, pmra, pmdec and the number of sigma at which the proper motion of the candidate differ from the host star's proper motion. The candidates are rejected at $3\sigma$ or higher.

\section{Analysis and discussion}\label{discussion}

\subsection{Sensitivity and completeness}\label{ccurves} 
\begin{figure}[h]
\centering
\epsscale{1}
\includegraphics[width=\linewidth]{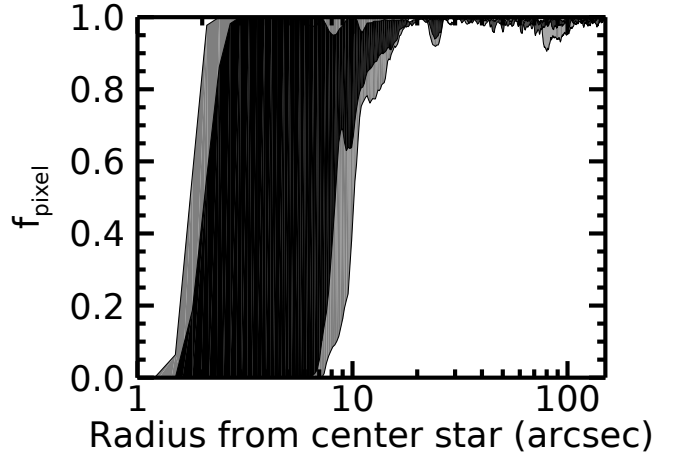}
\caption{\label{fracpix} Fraction of clean pixels where a companion could be detected as a function of the separation from the target star in the $J$-band images. 50\% of the stars have a fraction of pixel that is included in the black area while the grey area represents 80\% of the stars. For most stars, the fraction of clean pixels reaches 90\% at 10\arcsec.   }
 \end{figure} 
 
\begin{figure}[h]
\centering
\epsscale{1}
\includegraphics[width=\linewidth]{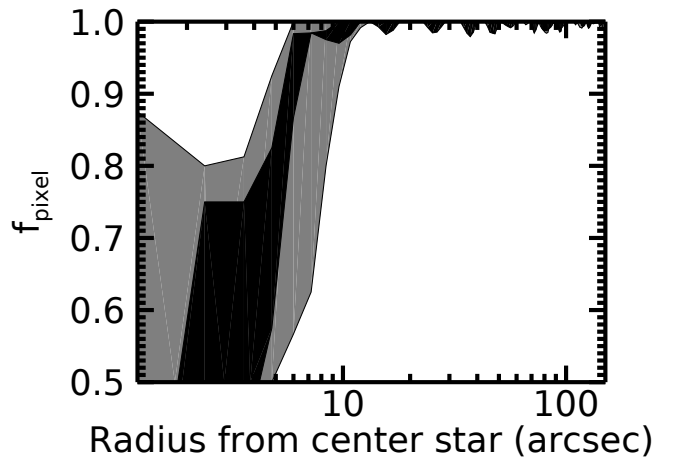}
\caption{\label{fracpix_spitzer} Same as \ref{fracpix} for $Spitzer$/IRAC observations at $[4.5]$. For most stars, the fraction of clean pixels reaches 98\% at 10\arcsec. }
 \end{figure} 
 

For each image of our survey, the sensitivity to planets of a given semi-major axis and mass can be determined using the limiting magnitude reached as a function of the projected separations from the star and the corresponding fraction of pixels where a companion could have been detected. In computing these detection completeness maps for all stars in our sample, we adopted an approach similar to that of \cite{nielsen_constraints_2008} and \cite{naud_psym-wide:_2017}, relying on a Monte Carlo simulation. 

First, for a given image and a given separation from the star, the fraction of clean pixels, i.e., pixels where a companion could have been detected if indeed it was present, was simply determined by counting pixels at that separation that were not flagged as bad, not saturated, and not affected by the presence of a star. Figures~\ref{fracpix} and \ref{fracpix_spitzer} show this fraction as a function of separation from the star for the $J$-band images and the $[4.5]$-band images, respectively. In most cases, at 10\arcsec\ the fraction reaches 0.9 for the $J$-band images and 0.98 for the $[4.5]$-band images. In a few cases, the target star is in the galactic plane, making the detection of a companion harder and the fraction lower. Huge variations in f$_{\rm pixel}$ at smaller separations come from the different magnitude of the central stars, and the associated different areas affected by saturation. Some  stars of the sample are very saturated and thus f$_{\rm pixel}$ is very low at small separation while the M dwarfs of our sample are not saturated and thus a higher f$_{\rm pixel}$ is reached at smaller separations.  In general, the fraction of pixel for an individual target can be fitted by a logistic function with the shape of $1/(e^{-a_0(x-a_1)}+e^{a_2})$, where $a_0$ is the steepness of the curve, $a_1$ is the x-values of the mid point and $a_2$ is typically close to 0. Table~\ref{fitsig} and Table~\ref{fitsigspitzer} show the values of the 3 parameters for each target of the sample for the $J$-band and [4.5] images respectively ($a_0$ varies from -10 to 40,  $a_1$ goes from 0 to 14, and $a_2$ is close to 0). 

Next, we defined a grid of masses and semi-major axes, with the masses equally spaced in logarithmic scale between 0.5 and 15~\mj\ and the semi-major axes equally spaced in logarithmic scale between 100 and 5000~AU. For each point of the grid, we simulated $10^4$ planets. Each planet has an eccentricity taken randomly from the eccentricity distribution reported in \cite{kipping_parametrizing_2013}, which is taken from the eccentricity from RV planets. Then we used the method of \cite{brandeker_deficit_2006} and \cite{brandt_statistical_2014} to find the instantaneous projected separation of each planet, given their eccentricity, semi-major axis, and some random inclination and time of observation. The projected separation in AU was finally converted to a projected angular separation in arcsec by dividing by the star distance, which is sampled uniformly within its interval of uncertainty.

For each grid point, we converted the mass into a $J$-band absolute magnitude using the AMES.Cond evolution models \citep{baraffe_evolutionary_2003}  and the ages of the targets from Table~\ref{asso} and Table~\ref{listeasso}. We randomly sampled the age of each generated planet uniformly between the uncertainties given for the appropriate moving group (see Table~\ref{asso}).  We then used the known distance of the star to convert the planets' absolute magnitudes to apparent magnitudes, and compared these to the detection limits found earlier to assess the detectability of each planet. If a planet was brighter than the detection limit, we used the fraction of clean pixels found earlier at that separation as the detection probability; otherwise the planet was assigned a detection probability of zero. This was repeated for each simulated planet, and the results were averaged to find the probability of detection at each point of the grid. 
 
\begin{figure*}[ht]
\centering
\epsscale{1}
\plottwo{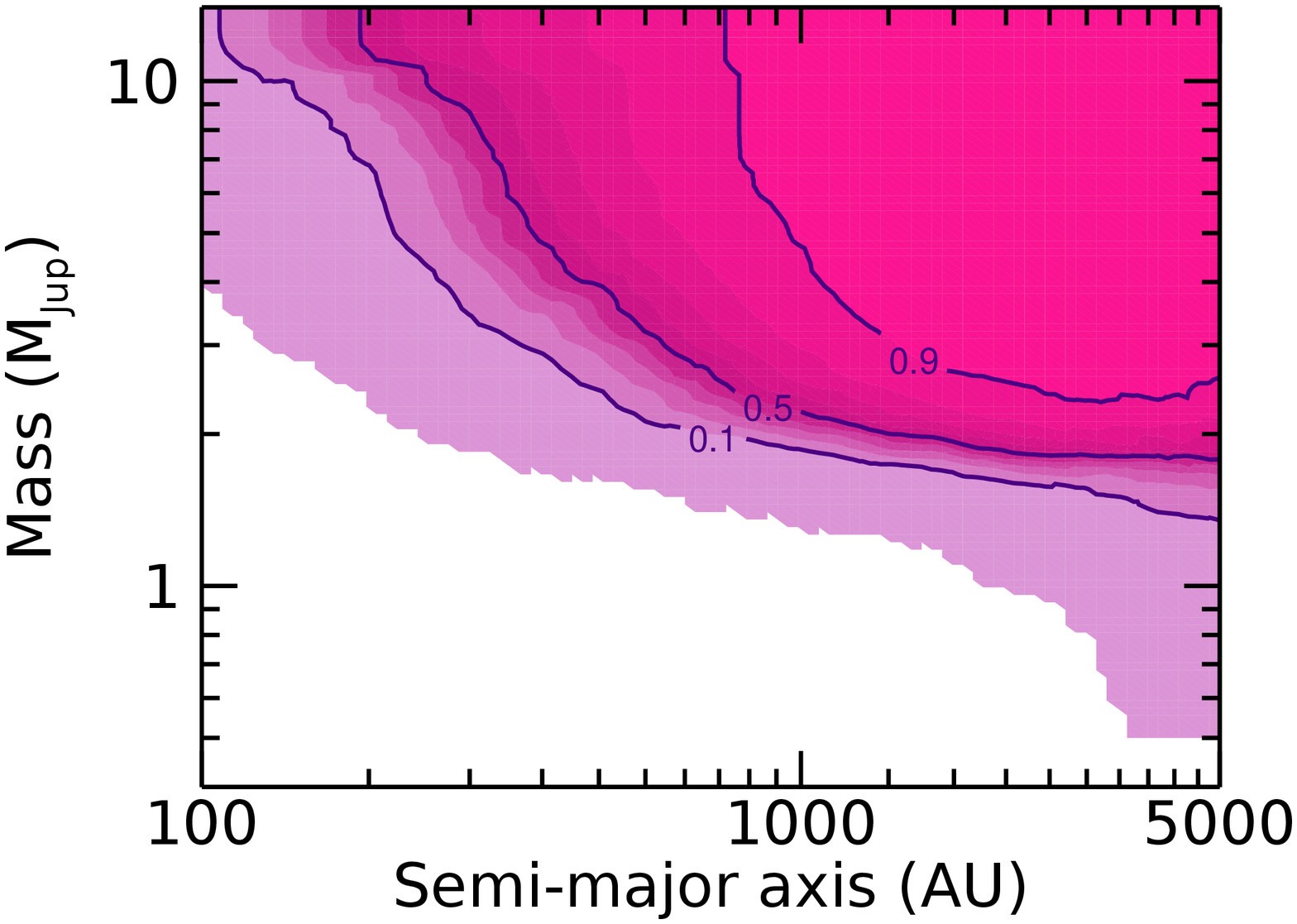}{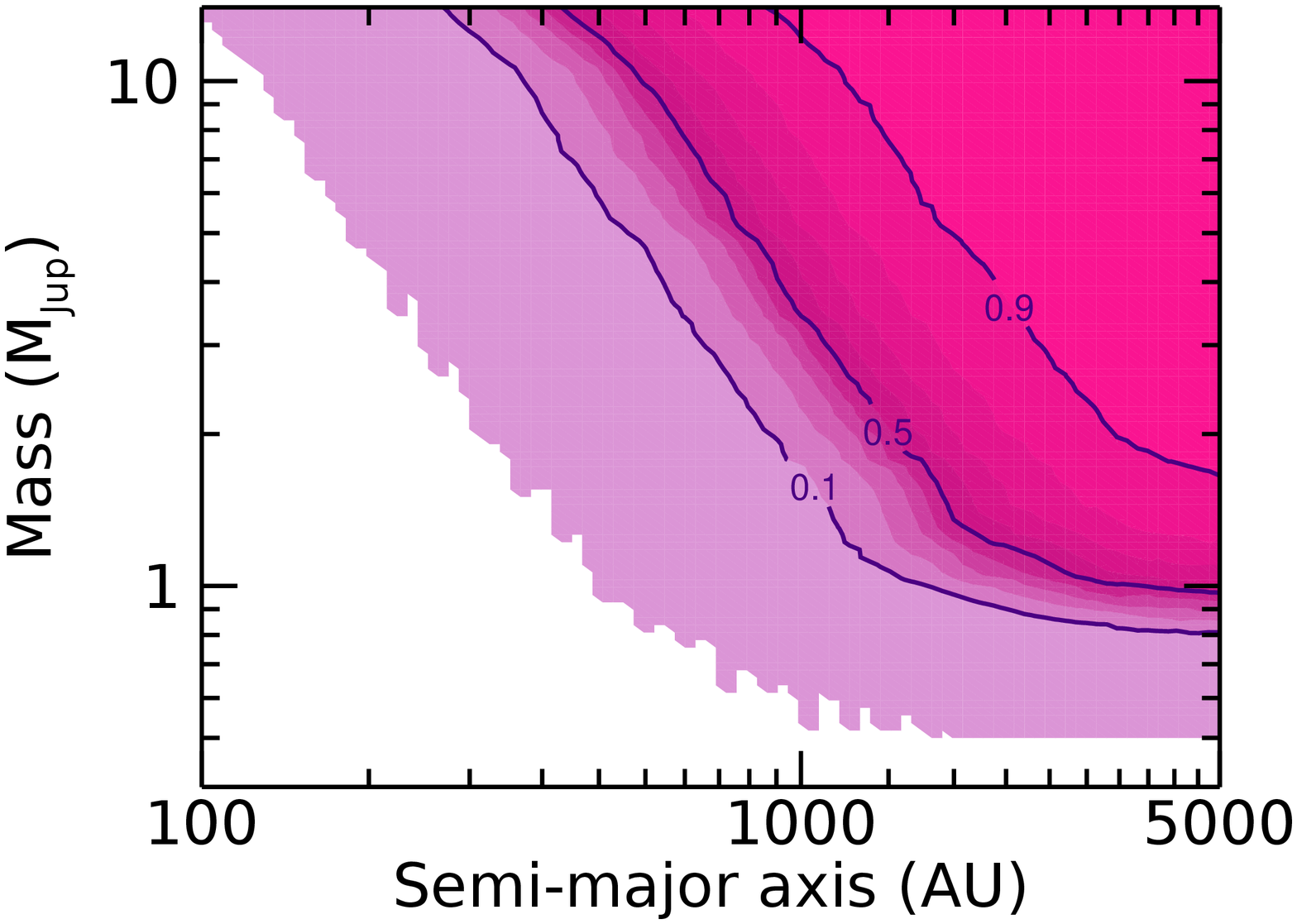}
\caption{\label{cc} Completeness map for the $J$-band images on the left and for the [4.5] images on the right. They show the probability of detecting a planet with a mass between 1 and 13~\mj\ as a function of the separation from the host star. Curves for 10\%, 50\% and 90\% are shown.  }
\end{figure*}

This procedure was repeated for all targets of the sample. The sensitivity of the whole survey was calculated by taking the median of all the detection probability maps. Two completeness maps were made this way, one for the $J$-band images (Figure~\ref{cc}, left) and one for the $[4.5]$-band images  (Figure~\ref{cc}, right). The ground-based survey is mostly sensitive to objects with masses higher than 2~\mj\ with a semi-major axis of more than 1000~AU while the $Spitzer$ survey is sensitive to planets slightly less massive (down to 1~\mj) at larger separations. 

\begin{figure*}[ht]
\centering
\plotone{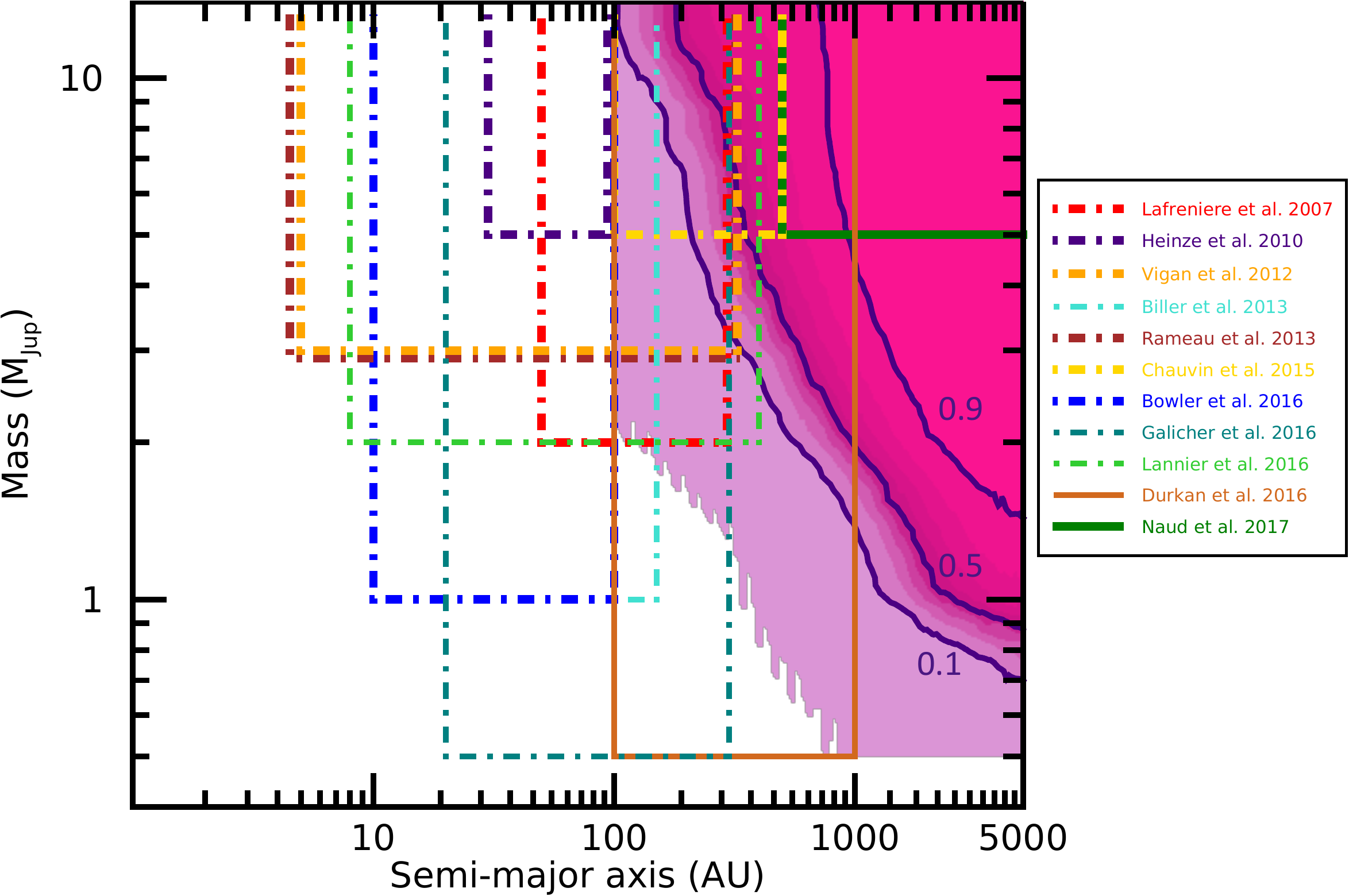}
\caption{\label{comparelitt} Overall completeness map for our survey. Our results are shown in shades of magenta and the contours correspond to the probability of detecting a planet of a giving mass and semi-major axis.
The solid green box is the PSYM-WIDE survey \citep{naud_psym-wide:_2017},  the solid brown box is the survey of \cite{durkan_high_2016}, and the dashed-dotted boxes correspond to high contrast direct imaging surveys: PALMS in blue \citep{bowler_imaging_2016}, GPDS in red \citep{lafreniere_gemini_2007}, NaCo Survey of Young Nearby Dusty Stars \citep{rameau_survey_2013} in brown, NaCo-LP in yellow \citep{chauvin_vlt/naco_2015}, IDPS-AF in orange \citep{vigan_international_2012}, MMT L$^\prime$ and M-band Survey of Nearby Sun-like Stars \citep{heinze_constraints_2010} in purple, Gemini NICI Planet-finding Campaign \citep{biller_gemini/nici_2013} in turquoise, MASSIVE in lime green \citep{lannier_massive:_2016} and IDPS in olive green \citep{galicher_international_2016}.  Our observations probe larger semi-major axes than AO imaging surveys, but are insensitive to semi-major axes where AO observations are mostly sensitive. }
 \end{figure*}

\begin{figure}[ht]
\centering
\includegraphics[width=\linewidth]{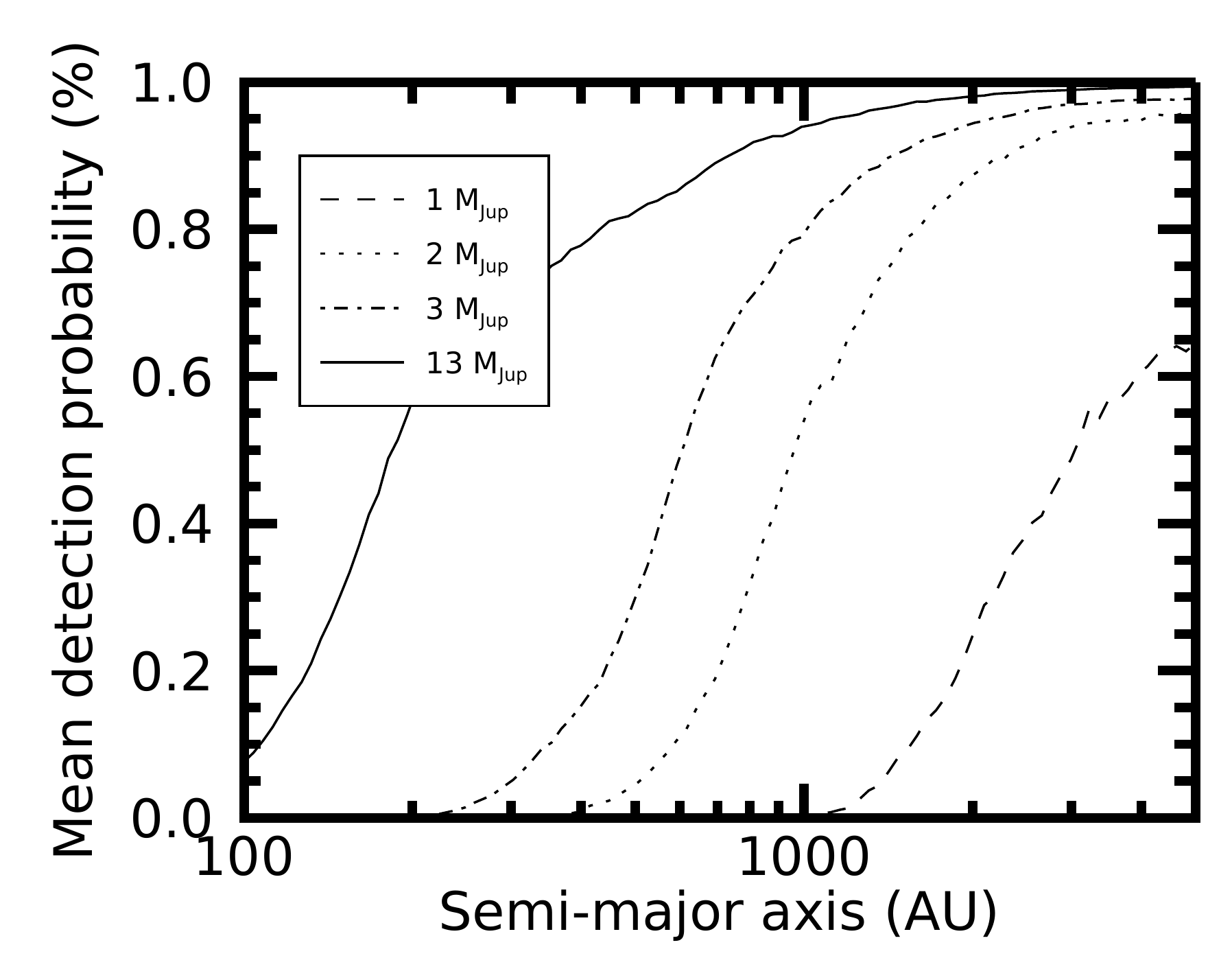}
\caption{\label{ccurvesboth} Mean detection probability for 1~\mj\ (dash), 2~\mj\ (dot), 3~\mj\ (dash-dot) and 13~\mj\ (solid) companions as a function of the semi-major axis in AU.    }
 \end{figure}

The completeness maps for each star of the sample and for both $J$ and $[4.5]$ bands were combined to build the overall completeness map of the survey. For each star at each point of the grid, the highest probability was taken between the completeness map of the $J$-band images and the [4.5] images. The two-band combined completeness maps were then averaged over all stars to obtain the overall survey completeness maps, see Figure~\ref{comparelitt}. Figure~\ref{ccurvesboth} shows the mean detection probability as a function of semi-major axis for planetary objects with masses of 1~$M_{\rm Jup}$, 2~$M_{\rm Jup}$, 3~$M_{\rm Jup}$ and 13~$M_{\rm Jup}$, taken from the overall completeness map of the survey. The maximal probabilities of detection are respectively 64\%, 95\%, 98\%, and 99\%. Our survey is mostly sensitive to planets with masses of 2\mj\ and above, as the detection probability falls very rapidly between 2 and 1~$M_{\rm Jup}$. 
 
Our results probe an area of the semi-major axis--mass diagram that has not been studied before. Figure~\ref{comparelitt} shows our completeness map compared to the regions probed by the following other studies:
the PSYM-WIDE survey \citep{naud_psym-wide:_2017}, aiming at discovering planetary mass objects on wide orbits around K5-L5 dwarfs, the PALMS survey \citep{bowler_imaging_2016}, a deep coronagraphic study of 78 single young nearby (<40~pc) M dwarfs,  the GPDS survey \citep{lafreniere_gemini_2007}, a survey of young stars searching for giant planets on large orbits, the NaCo Survey of Young Nearby Dusty Stars \citep{rameau_survey_2013}, which targeted 59 young nearby AFGK stars, the NaCo-LP survey \citep{chauvin_vlt/naco_2015}, which focused on 86 young, bright, and primarily FGK stars, the IDPS-AF survey \citep{vigan_international_2012}, which observed 42 AF stars, the MMT L$^\prime$ and M-band Survey of 54 nearby FGK stars  \citep{heinze_constraints_2010}, the Gemini NICI Planet-finding Campaign \citep{biller_gemini/nici_2013}, which targeted 230 young stars of all spectral types, MASSIVE \citep{lannier_massive:_2016}, which targeted  58 young and nearby M-type dwarfs, the IDPS survey \cite{galicher_international_2016}, which combine results for 292 young nearby stars and \cite{durkan_high_2016} who studied 121 nearby stars observed with $SPITZER$/IRAC.
On the whole, this survey is a good complement to AO imaging surveys, being mostly sensitive at separations of several hundreds of AU but insensitive at semi-major axes of less than $\sim$ 150~AU, where AO imaging surveys are most sensitive.

 

 

\subsection{Constraints on additional companions in systems with known directly imaged companions}\label{ccurves}

\begin{figure*}[t!]
$\begin{array}{rl}
    \includegraphics[width=0.5\textwidth]{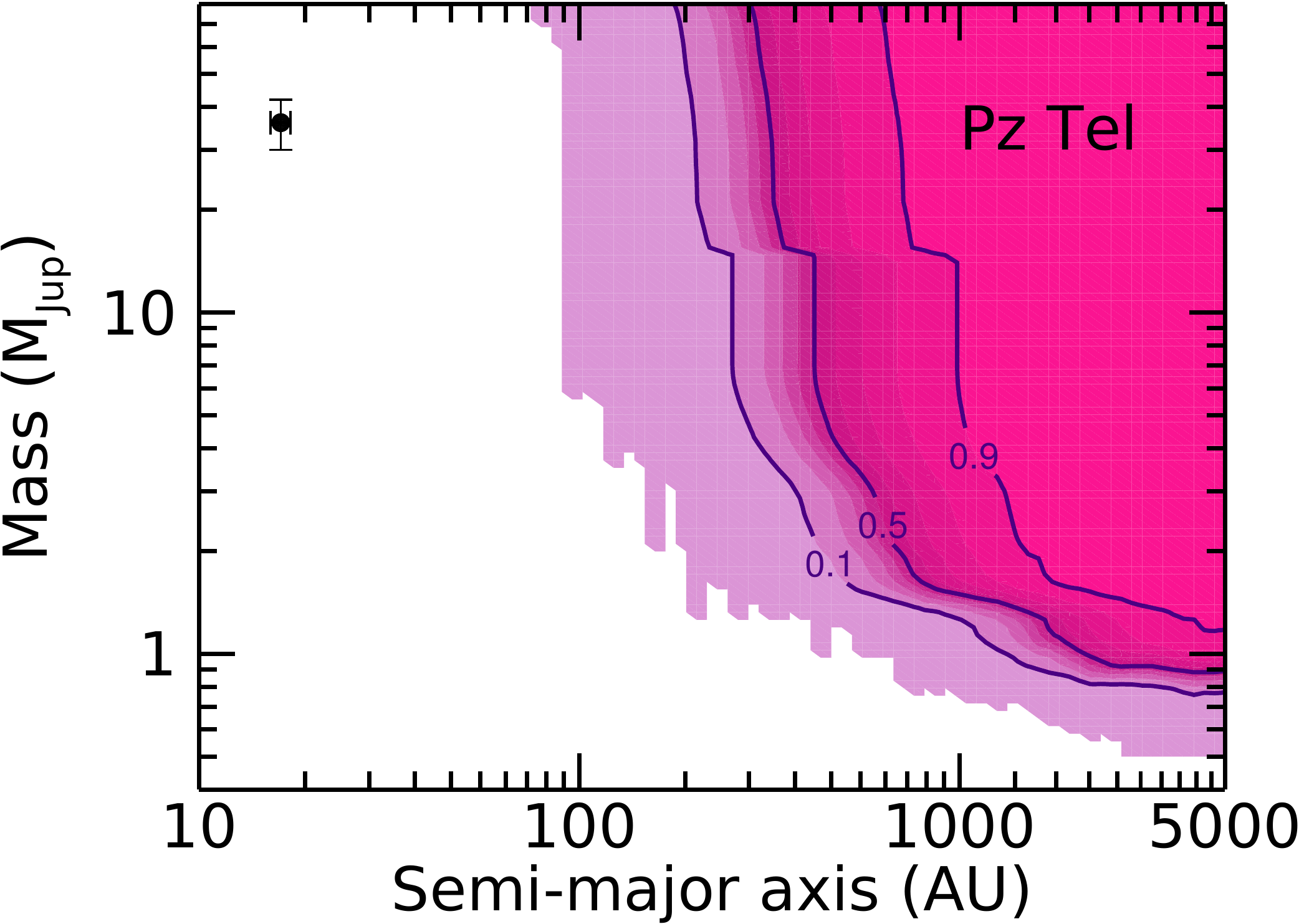} &
    \includegraphics[width=0.5\textwidth]{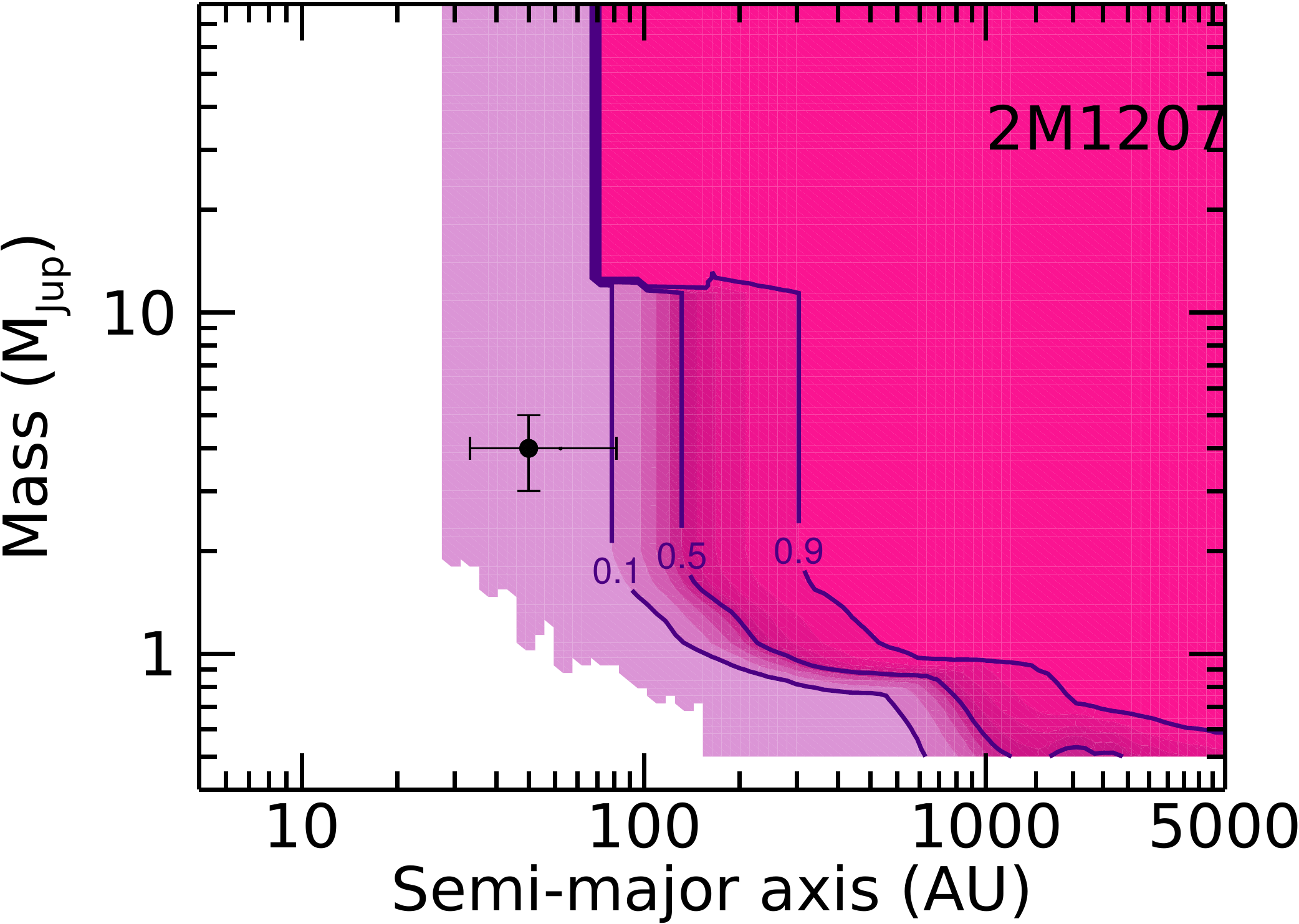}\\
    \includegraphics[width=0.5\textwidth]{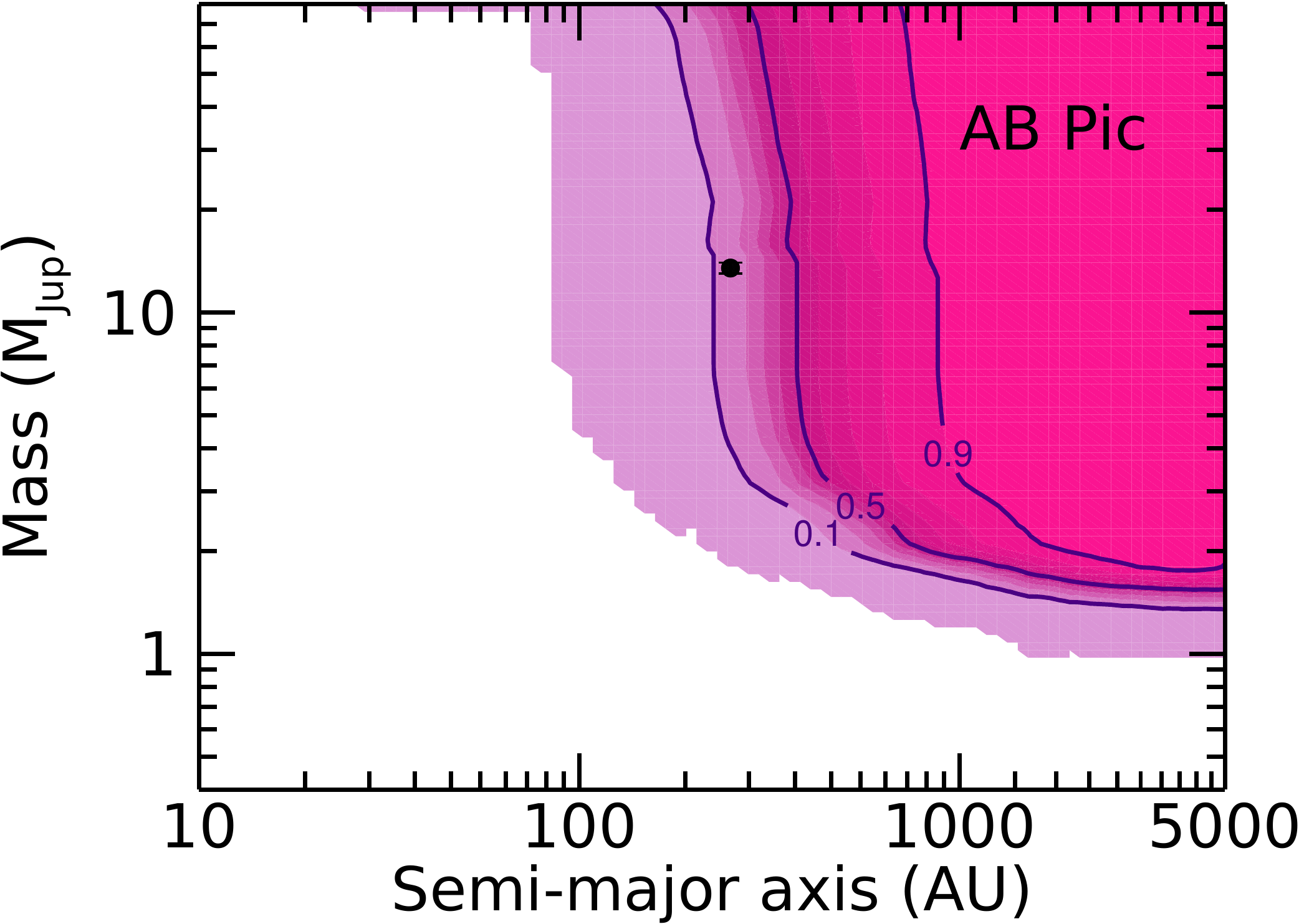} &
    \includegraphics[width=0.5\textwidth]{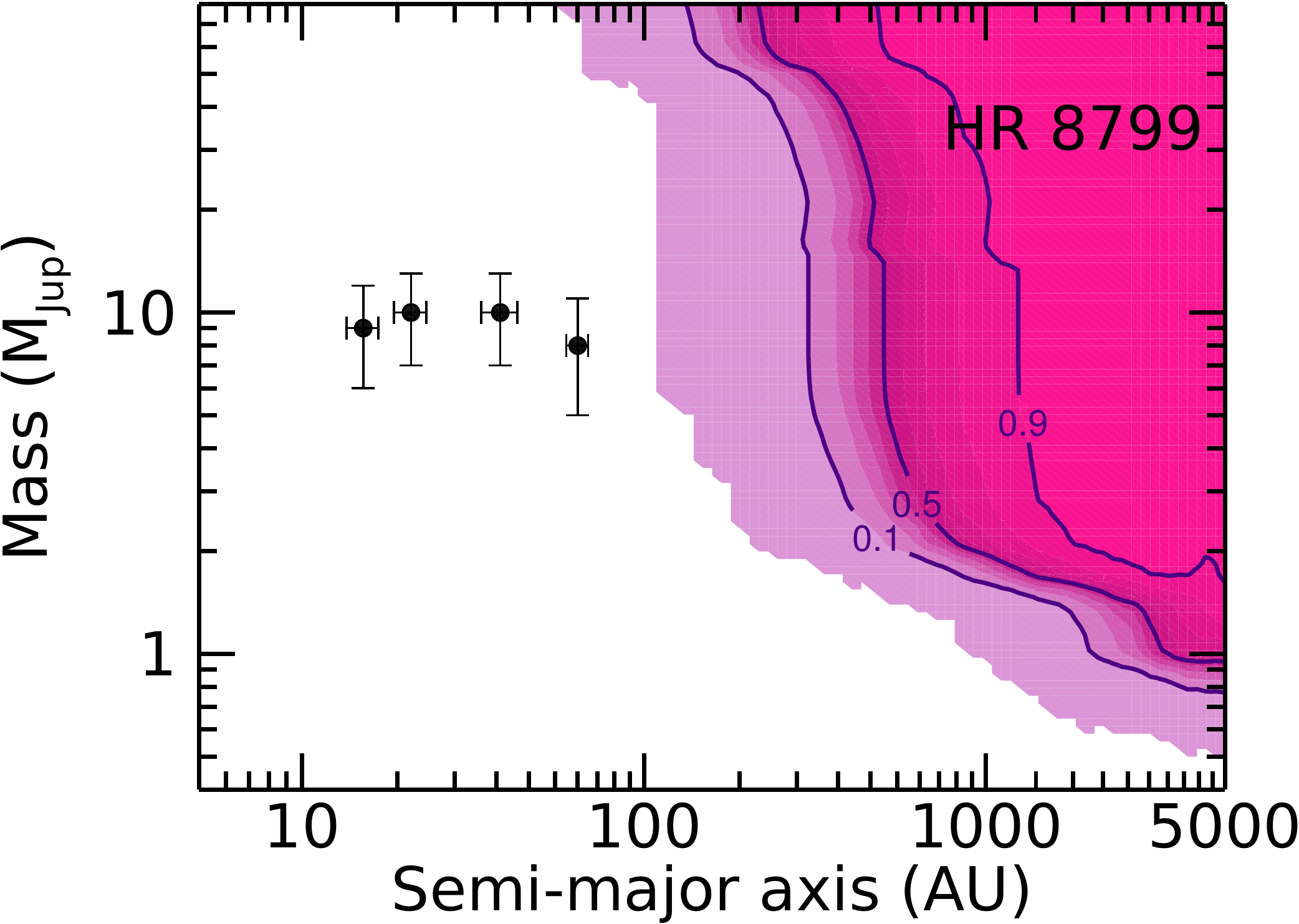}\\
    \includegraphics[width=0.5\textwidth]{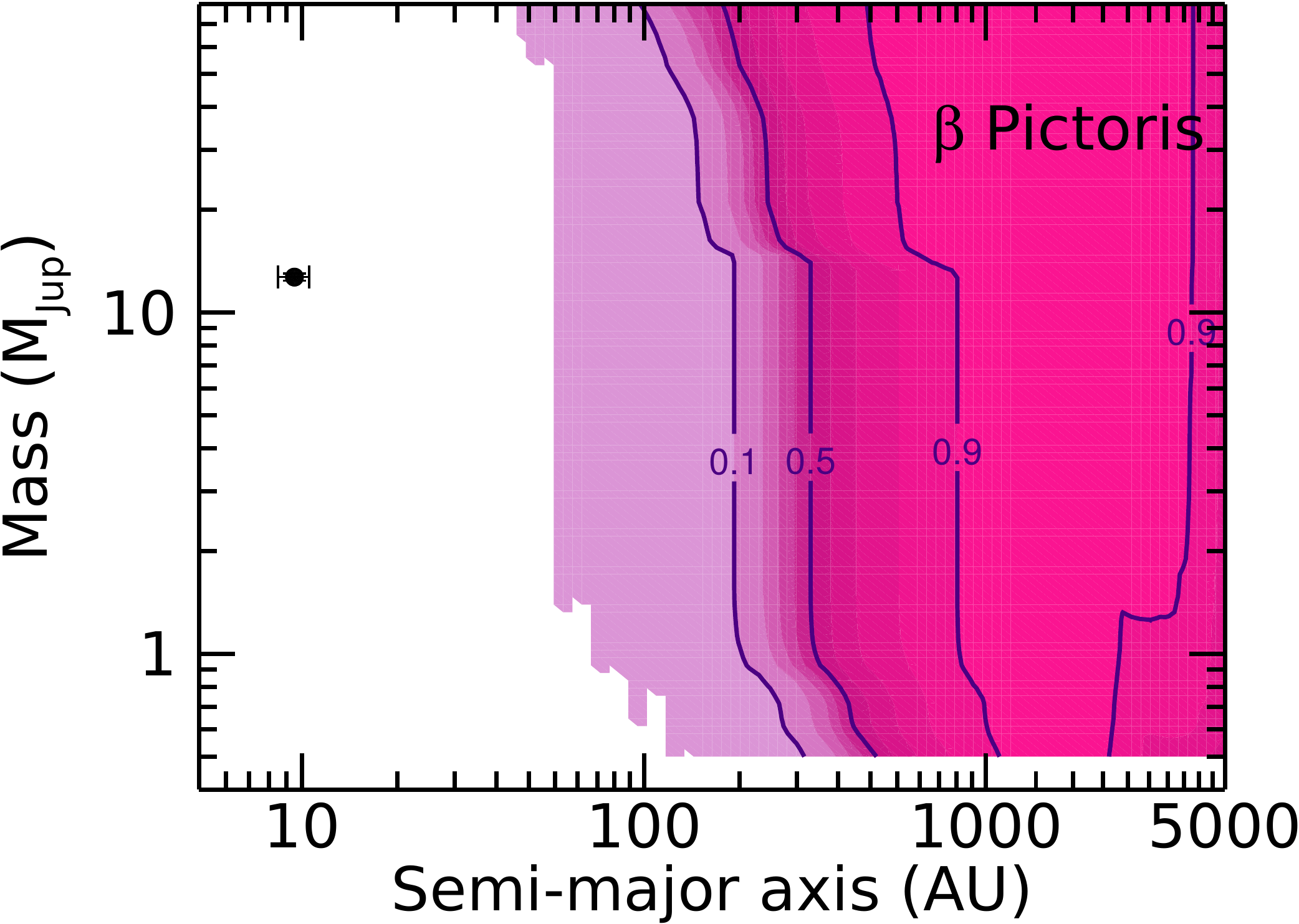} &
    \includegraphics[width=0.5\textwidth]{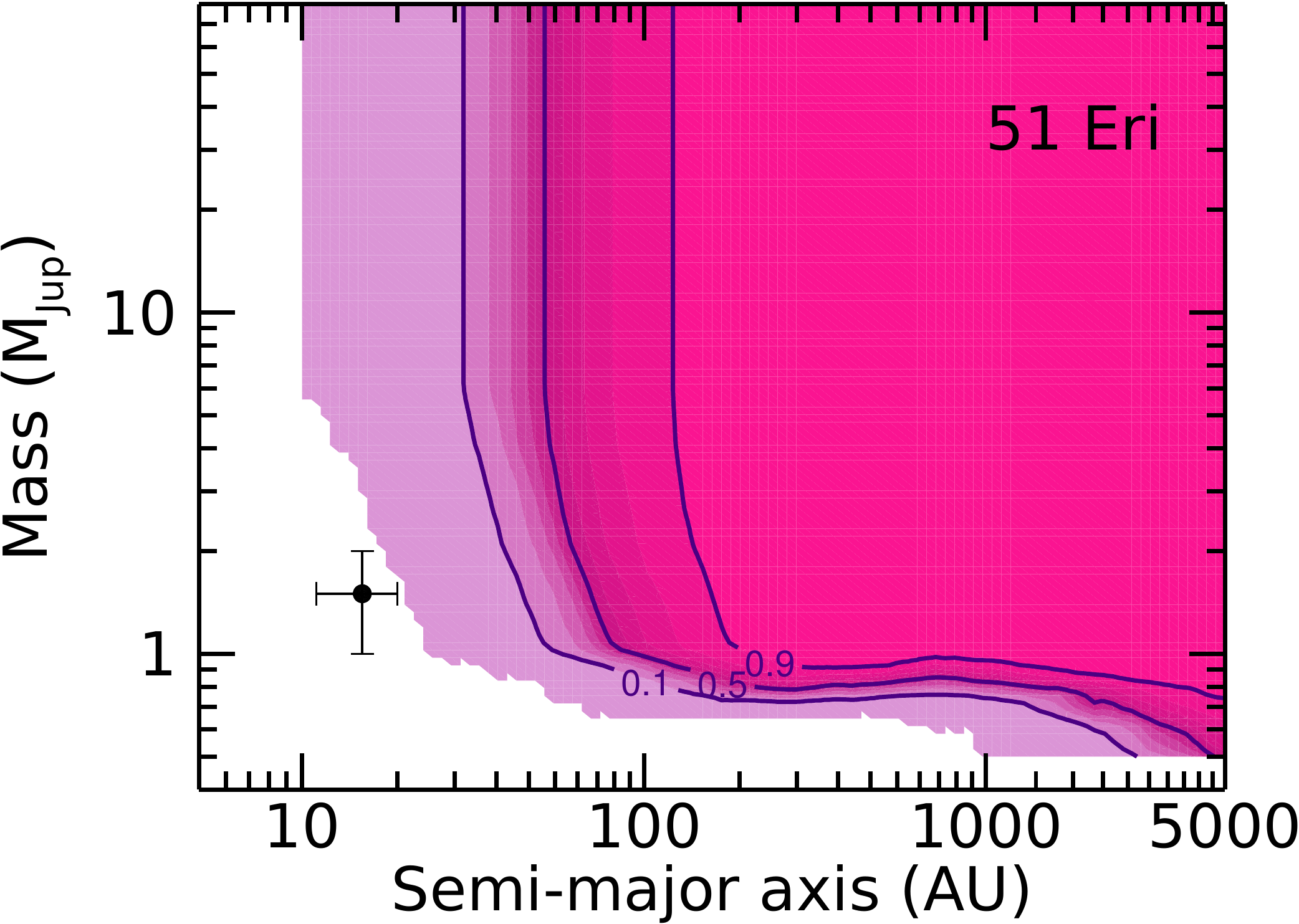}\\
    \end{array}$
\caption{\label{cc_all} Contrast curves for Pz Tel, 2M1207, AB~Pic, HR~8799, $\beta$~Pictoris and 51~Eri. Known companions are shown as black points with error bars, using masses from hot start models. See text for references for the masses.}
\end{figure*}

At least one planetary mass or brown dwarf companion was previously found around 6 stars in our sample; most of these companions were found using high-contrast AO imaging. Our search, being sensitive to much wider separations and reaching lower masses, adds valuable constraints on the presence of additional companions in these systems. We provide in Figure~\ref{cc_all} the individual completeness maps from our survey for these six systems.

The companion Pz~Tel~B, a $36 \pm 6$\mj\ brown dwarf orbiting at 16.4$\pm$1 AU from a pre-main sequence G9 star member of the $\beta$-Pictoris association, was found by 
\cite{biller_gemini_2010} using VLT/NACO. We put constraints on the presence of companions at larger orbits (see Figure~\ref{cc_all}, top left). At a confidence level of more than 90\%, we can reject a companion with masses as low as 1--2~\mj\ at 2000--5000~AU.

The companion 2M1207 b, a 4$\pm$1~\mj\ object \citep{chauvin_giant_2004} orbiting at 46 $^{+37}_{-15}$ AU \citep{blunt_orbits_2017} around the young brown dwarf TWA27, member of the TW Hydrae association at 52 pc, was discovered using VTL/NACO. Our survey put strong constraints on the presence of $>$ 10 \mj\ objects in the system, as they should have been detected at separations from 100 to 5000 AU. Moreover, at a distance of 1000 AU, the detection probability of 1 \mj\ object is about 80\%. Our survey covers quite well the regime of separations $>$ 1000 AU and masses $>$ 1~\mj\  (see Figure~\ref{cc_all}, top right). No companion was detected by our survey. 

\cite{chauvin_companion_2005} found a 13.5 $\pm$ 0.5~\mj\ object at 250 AU of AB Pic, a K2V star member of the Tucana-Horologium association, by using VLT/NACO. Figure~\ref{cc_all}, middle left, presents the completeness reached by our survey. We put strong constraints on the presence of companions of 2\mj\ or more at separations higher than 1000 AU. 

\cite{marois_direct_2008,marois_images_2010} used AO observations with Keck/NIRC2 and Gemini/NIRI to find 4 planets of 7$^{+4}_{-2}$, 10$^{+3}_{-3}$, 10$^{+3}_{-3}$ and 9$^{+4}_{-4}$ \mj\ at respectively $\sim$68, 43, 27 and 17 AU from HR~8799\citep{wertz_vlt/sphere_2017}, an A5V star member of the Columba association. We probed a region in mass that is equivalent to the planets already known, but at much larger semi-major axes. We put good constraints on the presence of companion with $\geq$ 4 \mj\ and semi-major axis greater than 1500 AU. 

\cite{lagrange_probable_2009} found a 12.7$\pm0.3$ \mj\ \citep{morzinski_magellan_2015} planet at 9.2$^{+1.5}_{0.4}$~AU \citep{millar-blanchaer_beta_2015} orbiting $\beta$ Pictoris, an A6V star member of the $\beta$ Pictoris association, using high-contrast VLT/NACO observations. Our observations put strong constraints on the existence of objects of 1~\mj\ or more at semi-major axes of >1000~AU. 

A 1--2~\mj\ \footnote{This mass was inferred from hot start model from \citep{marley_luminosity_2007}. It is also possible that the mass is anywhere between 2--12~\mj\ according to the cold start model from \cite{fortney_synthetic_2008}.} \citep{rajan_characterizing_2017} object orbiting 51~Eri at $\sim$14~AU, an F0IV star, was found by \cite{macintosh_discovery_2015} using Gemini/GPI. 51~Eri is part of a triple system, bound to and separated by $\sim$2000~AU from GJ3305AB, an M+M binary of unresolved spectral type M0 \citep{montet_dynamical_2015}. Our survey put strong constraints on the presence of companions of mass $>$ 1~\mj\ at semi-majors axes between 100 and 5000 AU. 

\subsection{Planet frequency}\label{planet_freq}

Based on the null result of our survey, and our completeness limits calculated in section~\ref{ccurves}, we evaluated an upper limit to the frequency of occurrence of planets at large semi-major axis (1000--5000~AU), following the method developed by \cite{lafreniere_gemini_2007}. 

If we have N=177 stars enumerated from j=1 to N, and we survey an interval of mass going from 1 to 13~\mj\ and an interval of semi-major axis of 1000 to 5000 AU, then we define $f$ to be the fraction of stars with at least one companion in the intervals and $p_j$ the probability of detecting such a companion. This probability is computed from the completeness map calculated previously by taking the mean of the probability at each point of the 100$\times$100 grid. Since the grid is uniform in logarithmic space, this amounts to assuming that the semi-major axis and the mass are distributed uniformly in log. The detections in the survey are characterized by the set $\{d_j\}$, and in our case, since the survey gave a null result (all known companions around our targets were too close-in to be seen in our data), $d_j$=0 for all $j$. The probability of observing the set $\{d_j\}$ in our survey is given by the following binomial likelihood, 

\begin{equation}
\mathcal{L}(\{d_j\}|f) = \prod_{j=0}^{N} (1-fp_j)^{1-d_j}(fp_j)^{d_j} \,.
\end{equation}

Then according to Bayes theorem, the posterior distribution for $f$, in light of our results, is given by, 
\begin{equation}
p(f|\{d_j\}) = \frac{\mathcal{L}(\{d_j\}|f)p(f)}{\int_{0}^{1} \mathcal{L}(\{d_j\}|f)p(f) \rm{d}f} \,,
\end{equation}
where $p(f)$ is the prior probability on $f$, reflecting our state of knowledge independently of our new data. One has to be careful in the choice of the prior, and here we elected to use a non-informative Jeffrey's prior \citep[see][]{berger_formal_2009} , given by,
\begin{equation}
P(f) = \frac{1}{\pi}\frac{1}{\sqrt{f}}\frac{1}{\sqrt{1-f}} \,.
\end{equation}

For our survey with no detection, the posterior distribution of $f$ peaks at 0, and we can only set an upper limit on $f$ (by integrating the posterior from 0 to the fraction $f$ that give a probability matching the desired confidence level).

We obtained an upper limit for the fraction of stars with at least one planet of $f_{max} = 0.03$ at a 95\% confidence level, for planets with masses between 1 and 13~\mj\ and semi-major axis between 1000 and 5000~AU distributed uniformly in log space.

\section{Conclusions}
A sample of 177 young stars, bona fide members of moving groups, were observed between 2014B and 2017B by CFHT's MegaCam in the $z_{ab}^\prime$-band and WIRCam in the $J$-band, or Gemini GMOS-S in the $z_{ab}^\prime$-band and Flamingos-2 in the $J$-band, as well as with $Spitzer$/IRAC at [3.6] and [4.5] to search for planetary mass companions on very wide orbits (up to 5000 AU). The survey made use of the very red $z^\prime-J$ and [3.6]-[4.5] colors intrinsic to such objects and reached good sensitivities down to objects of 1~$M_{\rm Jup}$. Four candidates were identified through colors selection but proper motion follow-up obtained a year after the first epoch rejected the candidates. No planet was found. This null result allowed us to set an upper limit of 0.03 for the fraction of stars with at least one planet with mass between 1 and 13~\mj\ and semi-major axis between 1000 and 5000 AU, at a 95\% confidence level, assuming logarithmically uniform distributions in planet mass and semi-major axis.While it was not the main objective of the survey, our data also constrain the frequency of brown dwarfs to be less than 2.2\% for objects with masses between 13 and 80\mj\ and for semi-major axis between 1000 and 5000 AU.

As mentioned above, the formation process by which Jupiter-like objects on wide orbits form has been the subject of an ongoing debate. The very low occurrence rate for planets at 1000-5000~AU found by our survey indicates that neither core accretion nor disk instability is actually efficient at forming gas giants at these large separations. It is possible that the few known instances of planets at such large separations from their host star represent the low-mass tail end of distribution of brown dwarf companions that form like stars, rather than objects that form like planets. More quantitative implications of our results on the properties of the overall distribution of planets around stars, as well as on the formation mechanism of very distant companions will be explored in a forthcoming paper, where we will further incorporate the results of AO surveys.

\acknowledgments
Based on observations obtained at the Gemini Observatory through programs number GS-2014B-Q-2, GS-2015A-Q-71, GS-2015B-Q-57, GS-2016A-Q-69, GS-2016B-Q-33, GS-2017A-Q-58 and GS-2017B-Q-34. The Gemini Observatory is operated by the Association of Universities for Research in Astronomy, Inc., under a cooperative agreement with the National Science Foundation (NSF) on behalf of the Gemini partnership: the NSF (United States), the National Research Council (Canada), CONICYT (Chile), the Australian Research Council (Australia), Ministério da Ciência, Tecnologia e Inovação (Brazil), and Ministerio de Ciencia, Tecnología e Innovación Productiva (Argentina).

Based on observations obtained with MegaPrime/MegaCam, a joint project of CFHT and CEA/DAPNIA, at the Canada-France-Hawaii Telescope (CFHT) which is operated by the National Research Council (NRC) of Canada, the Institut National des Science de l'Univers of the Centre National de la Recherche Scientifique (CNRS) of France, and the University of Hawaii. 

The Pan-STARRS1 Surveys (PS1) have been made possible through contributions of the Institute for Astronomy, the University of Hawaii, the Pan-STARRS Project Office, the Max-Planck Society and its participating institutes, the Max Planck Institute for Astronomy, Heidelberg and the Max Planck Institute for Extraterrestrial Physics, Garching, The Johns Hopkins University, Durham University, the University of Edinburgh, Queen's University Belfast, the Harvard-Smithsonian Center for Astrophysics, the Las Cumbres Observatory Global Telescope Network Incorporated, the National Central University of Taiwan, the Space Telescope Science Institute, the National Aeronautics and Space Administration under Grant No. NNX08AR22G issued through the Planetary Science Division of the NASA Science Mission Directorate, the National Science Foundation under Grant No. AST-1238877, the University of Maryland, and Eotvos Lorand University (ELTE) and the Los Alamos National Laboratory.

This work has made use of data from the European Space Agency (ESA) mission {\it Gaia} (\url{https://www.cosmos.esa.int/gaia}), processed by
the {\it Gaia} Data Processing and Analysis Consortium (DPAC,
\url{https://www.cosmos.esa.int/web/gaia/dpac/consortium}). Funding
for the DPAC has been provided by national institutions, in particular
the institutions participating in the {\it Gaia} Multilateral Agreement.


\facility{Gemini-South (Flamingos-2,GMOS-S), CFHT (WIRCam, MegaCam), Spitzer (Irac)}
\software{SExtractor \citep{bertin_sextractor:_1996}, Scamp \citep{bertin_scamp:_2010}, Swarp \citep{bertin_swarp:_2010}, CFHT'S Elixir pipeline}


\clearpage 

\startlongtable
\begin{deluxetable*}{ccccccccc}
\tablecaption{Properties of the sample of bona fide members\label{listecomplete}}
\tablewidth{700pt}
\tabletypesize{\scriptsize}
\tablehead{
\colhead{Name} & \colhead{RA} &
\colhead{DEC} & \colhead{SpT} &
\colhead{J} & \colhead{H} &
\colhead{K} & \colhead{W1} & \colhead{W2} \\
\colhead{} & \colhead{(J2000.0)} & \colhead{(J2000.0)} & \colhead{} &
\colhead{} &
\colhead{}& \colhead{}& \colhead{} & \colhead{}
}
\startdata
HIP 490&00 05 52.54&-41 45 11.0&G0V&6.464$\pm$0.011&6.189$\pm$0.017&
6.117$\pm$0.013&6.043$\pm$0.053&6.053$\pm$0.023\\\hline
HIP 560&00 06 50.08&-23 06 27.1&F3V&5.451$\pm$0.017&5.331$\pm$0.045&
5.240$\pm$0.019&5.245$\pm$0.072&5.013$\pm$0.036\\\hline
HIP 1113&00 13 53.01&-74 41 17.8&G8V&7.406$\pm$0.013&7.087$\pm$0.025&
6.962$\pm$0.017&6.888$\pm$0.035&6.932$\pm$0.020\\\hline
HIP 1134&00 14 10.25&-07 11 56.8&F5V&6.402$\pm$0.015&6.170$\pm$0.035&
6.073$\pm$0.015&6.049$\pm$0.046&5.999$\pm$0.024\\\hline
HIP 1481&00 18 26.12&-63 28 39.0&F8V&6.462$\pm$0.007&6.248$\pm$0.033&
6.149$\pm$0.009&6.141$\pm$0.048&6.102$\pm$0.023\\\hline
HIP 1910 AB&00 24 08.98&-62 11 04.3&M0V&8.385$\pm$0.019&7.708$\pm$0.031&
7.494$\pm$0.015&7.354$\pm$0.026&7.306$\pm$0.019\\\hline
HIP 1993&00 25 14.66&-61 30 48.3&M0V&8.615$\pm$0.021&7.943$\pm$0.037&
7.749$\pm$0.021&7.606$\pm$0.025&7.594$\pm$0.020\\\hline
GJ 2006A&00 27 50.23&-32 33 06.4&M3.5V&8.882$\pm$0.032&8.236$\pm$0.038&
8.012$\pm$0.033&7.720$\pm$0.021&7.541$\pm$0.017\\
\enddata
\end{deluxetable*}
\clearpage 

\startlongtable
\begin{deluxetable*}{cccccc}
\tablecaption{Properties of the sample of bona fide members\label{listeasso}}
\tablewidth{700pt}
\tabletypesize{\scriptsize}
\tablehead{
\colhead{Name} & \colhead{$\mu_{\alpha}cos\delta$} & \colhead{$\mu_{\delta}$} & \colhead{Radial Velocities} &
\colhead{Distance} & \colhead{Association} \\
\colhead{} &\colhead{(mas/yr)} & \colhead{(mas/yr)} & \colhead{(km/s)} & \colhead{(pc)} & \colhead{}
}
\startdata
                  HIP 490   &             97.53$\pm$0.38   &            -76.27$\pm$0.44   &                1.5$\pm$1.2   &             39.38$\pm$0.91   &          Tucana-Horologium
                 \\\hline
                  HIP 560   &             97.81$\pm$0.42   &            -47.12$\pm$0.21   &                6.5$\pm$3.5   &             39.38$\pm$0.58   &           $\beta$-Pictoris
                 \\\hline
                 HIP 1113   &             83.53$\pm$0.78   &            -47.89$\pm$0.75   &                9.3$\pm$0.2   &             44.40$\pm$1.61   &          Tucana-Horologium
                 \\\hline
                 HIP 1134   &            102.79$\pm$0.78   &            -66.36$\pm$0.36   &               -2.2$\pm$1.2   &             47.14$\pm$1.42   &                    Columba
                 \\\hline
                 HIP 1481   &             89.37$\pm$0.48   &            -59.46$\pm$0.50   &                6.4$\pm$0.4   &             41.54$\pm$0.89   &          Tucana-Horologium
                 \\\hline
              HIP 1910 AB   &             90.91$\pm$2.37   &            -47.25$\pm$3.04   &                6.6$\pm$0.6   &             52.96$\pm$7.63   &          Tucana-Horologium
                 \\\hline
                 HIP 1993   &             87.76$\pm$2.14   &            -57.48$\pm$2.37   &                6.4$\pm$0.1   &             45.80$\pm$5.07   &          Tucana-Horologium
                 \\\hline
                 GJ 2006A   &            117.40$\pm$2.80   &            -29.30$\pm$8.10   &                8.7$\pm$0.2   &             32.20$\pm$1.05   &           $\beta$-Pictoris
                 \\
\enddata
\end{deluxetable*}

\clearpage 
\startlongtable
\begin{deluxetable*}{ccccccccc}
\tablecaption{$J$-band observations for all the target in the sample\label{zpointJ}}
\tablewidth{700pt}
\tabletypesize{\scriptsize}
\tablehead{
\colhead{Name} & \colhead{Filter} &
\colhead{Instrument} & \colhead{Obs. Date} &
\colhead{Nexp} & \colhead{Exposition Time} & \colhead{FWHM} &
\colhead{Catalog}&\colhead{Conditions} \\
\colhead{} & \colhead{} & \colhead{} & \colhead{} &
\colhead{} &\colhead{(s)} & \colhead{('')} & \colhead{}& \colhead{}
}
\startdata
HIP490&J\char`_G0802&F2&2014-07-25&9&612&0.72&median&photometric\\\hline
HIP560&J\char`_G0802&F2&2014-07-25&9&612&1.31&2MASS&photometric\\\hline
HIP1113&J\char`_G0802&F2&2016-10-10&14&630&0.67&2MASS&photometric\\\hline
HIP1134&J&WIRCam&2016-08-18&16&960&0.70&2MASS&patchy clouds\\\hline
HIP1481&J\char`_G0802&F2&2014-07-25&9&612&1.01&VISTA&patchy clouds\\\hline
HIP1910&J\char`_G0802&F2&2016-09-16&14&630&0.77&2MASS&photometric\\\hline
HIP1993&J\char`_G0802&F2&2015-09-02&14&630&0.89&2MASS&patchy clouds\\\hline
GJ2006A&J\char`_G0802&F2&2014-07-25&9&612&0.69&2MASS&photometric\\
\enddata
\end{deluxetable*}

\clearpage
\clearpage
\startlongtable
\begin{deluxetable*}{ccccccccc}
\tablecaption{$z_{ab}^\prime$-band observation for all the target in the sample\label{zpointz}}
\tablewidth{700pt}
\tabletypesize{\scriptsize}
\tablehead{
\colhead{Name} & \colhead{Filter} &
\colhead{Instrument} & \colhead{Obs. Date} &
\colhead{Nexp} & \colhead{Exposition Time} & \colhead{FWHM} &
\colhead{Catalog}&\colhead{Conditions} \\
\colhead{} & \colhead{} & \colhead{} & \colhead{} &
\colhead{} &\colhead{(s)} & \colhead{('')} & \colhead{}& \colhead{}
}
\startdata
HIP490&z\char`_G0328&GMOS-S&2014-09-09&8&1200&5.57&skymapper&patchy clouds
\\\hline
HIP560&z\char`_G0328&GMOS-S&2014-10-10&8&1200&0.93&panstarrs&patchy clouds
\\\hline
HIP1113&z\char`_G0328&GMOS-S&2015-09-02&8&520&0.76&skymapper&patchy clouds
\\\hline
HIP1134&z.MP9901&MegaCam&2016-09-07&4&1244&3.20&skymapper&photometric\\\hline
HIP1481&z\char`_G0328&GMOS-S&2014-07-20&9&1350&0.89&median&patchy clouds\\\hline
HIP1910&z\char`_G0328&GMOS-S&2015-08-03&9&585&0.98&skymapper&patchy clouds
\\\hline
HIP1993&z\char`_G0328&GMOS-S&2015-08-24&11&715&0.88&skymapper&photometric
\\\hline
GJ2006A&z\char`_G0328&GMOS-S&2016-06-10&8&520&0.99&skymapper&patchy clouds
\\
\enddata
\end{deluxetable*}

\clearpage
\startlongtable
\begin{deluxetable*}{cccccccc}
\tablecaption{7$\sigma$ detection limits in the $J$ band \label{liste_detect_lim}}
\tablewidth{700pt}
\tabletypesize{\scriptsize}
\tablehead{
\colhead{Name} &
\colhead{a$_{min}$} & \colhead{a$_{max}$} &
\colhead{a$_{min}$} & \colhead{a$_{min}$} &
\colhead{$J$} & \colhead{M$_{J}$} &
\colhead{Mass Limit} \\
\colhead{} & \colhead{$'$} & \colhead{$'$} &
\colhead{AU} & \colhead{AU} & \colhead{} &
\colhead{'}& \colhead{M$_Jup$}
}
\startdata
HIP490&32&180&1241&7096&22.1&19.1&1.0\\\hline
HIP560&46&180&1808&7096&20.7&17.7&1.7\\\hline
HIP1113&5&180&232&8001&21.0&17.7&1.7\\\hline
HIP1134&19&630&877&29717&21.8&18.5&1.3\\\hline
HIP1481&27&180&1137&7486&20.6&17.5&1.8\\\hline
HIP1910&22&180&1163&9543&21.5&17.9&1.6\\\hline
HIP1993&23&180&1055&8254&21.4&18.1&1.5\\\hline
GJ2006A&26&180&823&5802&21.3&18.8&1.2\\
\enddata
\end{deluxetable*}

\clearpage
\startlongtable
\begin{deluxetable*}{cccccccc}
\tablecaption{7$\sigma$ detection limits in the [4.5] band \label{liste_detect_lim_spit}}
\tablewidth{700pt}
\tabletypesize{\scriptsize}
\tablehead{
\colhead{Name} &
\colhead{a$_{min}$} & \colhead{a$_{max}$} &
\colhead{a$_{min}$} & \colhead{a$_{min}$} &
\colhead{$[4.5]$} & \colhead{M$_{[4.5]}$} &
\colhead{Mass Limit} \\
\colhead{} & \colhead{$'$} & \colhead{$'$} &
\colhead{AU} & \colhead{AU} & \colhead{} &
\colhead{'}& \colhead{\,$M_{\rm Jup}$ }
}
\startdata
HIP490&28&479&1087&18858&18.7&15.7&0.3\\\hline
HIP560&25&479&993&18858&18.4&15.4&0.4\\\hline
HIP1113&28&479&1226&21261&18.8&15.6&0.3\\\hline
HIP1134&24&479&1132&22574&18.2&14.8&0.7\\\hline
HIP1481&25&479&1047&19892&18.5&15.4&0.4\\\hline
HIP1910&26&479&1398&25360&18.7&15.0&0.6\\\hline
HIP1993&22&479&989&21933&18.8&15.5&0.4\\\hline
GJ2006A&20&479&657&15417&18.8&16.2&$<0.5$\\
\enddata
\end{deluxetable*}

\begin{deluxetable*}{ccccccccc}
\tablecaption{Properties of the candidates without MIPS data or detection \label{liste_extra}}
\tablewidth{700pt}
\tabletypesize{\scriptsize}
\tablehead{
\colhead{RA} &
\colhead{DEC} & \colhead{Host Star} &
\colhead{$z_{ab}^\prime$} & \colhead{$J'$} &
\colhead{$[3.6]$} & \colhead{$[4.5]$} &
\colhead{Separation} & \colhead{Distance}  \\
\colhead{(J2000.0)} & \colhead{(J2000.0)} & \colhead{(mag)} &
\colhead{(mag)} & \colhead{(mag)} & \colhead{(mag)} &
\colhead{'}& \colhead{pc}
}
\startdata
             02:46:00.708   &                05:35:00.82   &                   HIP12925   &                   $>$23.83   &                   $>$21.48   &             16.68$\pm$0.08   &             16.76$\pm$0.08   &          67.5130$\pm$0.019   &               54.3$\pm$3.0  \\
             02:50:00.567   &                27:16:00.52   &                   HIP13209   &                   $>$22.73   &                   $>$20.93   &             15.59$\pm$0.07   &             15.62$\pm$0.07   &          69.7032$\pm$0.021   &               50.7$\pm$0.4  \\
             04:48:00.751   &               -80:46:00.34   &                   HIP22295   &                   $>$24.75   &                   $>$21.71   &             18.42$\pm$0.08   &             18.11$\pm$0.08   &          107.158$\pm$0.188   &               61.0$\pm$1.8  \\
             05:01:00.177   &               -20:03:00.08   &                   HIP23362   &                   $>$24.04   &                   $>$20.94   &             18.62$\pm$0.08   &             17.83$\pm$0.08   &          57.8078$\pm$0.017   &               60.6$\pm$2.1  \\
             05:01:00.270   &               -20:01:00.43   &                   HIP23362   &                   $>$24.04   &                   $>$20.94   &             17.77$\pm$0.08   &             17.32$\pm$0.08   &          80.2475$\pm$0.022   &               60.6$\pm$2.1  \\
             05:20:00.536   &               -39:45:00.03   &                   HIP24947   &                   $>$22.15   &                   $>$20.00   &             18.02$\pm$0.08   &             17.71$\pm$0.08   &          97.8883$\pm$0.035   &               48.3$\pm$0.9  \\
             06:00:00.277   &               -44:53:00.56   &                   HIP28474   &                   $>$24.82   &                   $>$20.92   &             18.36$\pm$0.05   &             18.07$\pm$0.05   &          75.5571$\pm$0.029   &               52.5$\pm$1.6  \\
\enddata
\end{deluxetable*}

\begin{deluxetable*}{ccccccccccc}
\tablecaption{Properties of the candidates identified through color cuts\label{listecand}}
\tablewidth{700pt}
\tabletypesize{\scriptsize}
\tablehead{
\colhead{Host star} &
\colhead{RA} & \colhead{DEC} &
\colhead{$M_{z_{ab}^\prime}$} & \colhead{$M_{J}$} &
\colhead{$M_{[3.6]}$} & \colhead{$M_{[4.5]}$} &\colhead{Separation} &
\colhead{$\mu_{\alpha}cos\delta$}  & \colhead{$\mu_{\delta}$} &\colhead{Rejected at }\\
\colhead{(J2000.0)} & \colhead{(J2000.0)} & \colhead{} &
\colhead{(mag)} & \colhead{(mag)} & \colhead{(mag)} & \colhead{(mag)} &\colhead{(AU)} &
\colhead{(mas/yr)}& \colhead{(mas/yr)}& \colhead{(sigma)}
}
\startdata
              HIP 14913 A   &                  48.085635   &                 -44.426938   &             18.26$\pm$0.13   &             15.98$\pm$0.06   &             13.82$\pm$0.10   &             13.31$\pm$0.09   &               5559$\pm$101   &           -46.71$\pm$26.00   &          16.75$\pm$30.2700   &                          3  \\
                HIP 29964   &                  94.595099   &                 -72.054682   &             18.35$\pm$1.70   &             16.88$\pm$0.02   &             14.46$\pm$0.07   &             13.46$\pm$0.05   &               1670$\pm$116   &            27.52$\pm$17.48   &          12.81$\pm$22.5900   &                          3  \\
                HIP 79881   &                  244.61165   &                 -28.608996   &             18.35$\pm$1.60   &             16.36$\pm$0.09   &             14.05$\pm$0.06   &             13.31$\pm$0.05   &               6365$\pm$442   &              3.73$\pm$9.76   &          -6.13$\pm$9.53000   &                          8  \\
                HD 152555   &                  253.51768   &                 -4.3371192   &             18.83$\pm$0.07   &             16.46$\pm$0.12   &             14.08$\pm$0.12   &             13.65$\pm$0.11   &                4895$\pm$89   &           -10.10$\pm$13.25   &         -12.86$\pm$17.6900   &                          5  \\
\enddata
\end{deluxetable*}

\clearpage

\startlongtable
\begin{deluxetable*}{cccc}
\tablecaption{Parametrization of the $J$-band images fraction of pixel as a logistic function}\label{fitsig}
\tablewidth{700pt}
\tabletypesize{\scriptsize}
\tablehead{
\colhead{Name} &
\colhead{a$_{0}$} & \colhead{a$_{1}$} &
\colhead{a$_{2}$} \\
\colhead{} & \colhead{} & \colhead{} &
\colhead{} 
}
\startdata
HIP490&8.56&0.83&0.0331\\\hline
HIP560&13.15&1.09&0.0777\\\hline
HIP1113&25.65&5.59&0.0308\\\hline
HIP1134&-0.03&0.16&0.0131\\\hline
HIP1481&9.54&0.86&0.0308\\\hline
HIP1910&27.53&3.61&0.0190\\\hline
HIP1993&16.40&2.14&0.0159\\\hline
GJ2006A&15.08&2.11&0.0723\\
\enddata
\end{deluxetable*}

\clearpage

\startlongtable
\begin{deluxetable*}{cccc}
\tablecaption{Parametrization of $SPITZER$'s fraction of pixel as a logistic function}\label{fitsigspitzer}
\tablewidth{700pt}
\tabletypesize{\scriptsize}
\tablehead{
\colhead{Name} &
\colhead{a$_{0}$} & \colhead{a$_{1}$} &
\colhead{a$_{2}$} \\
\colhead{} & \colhead{} & \colhead{} &
\colhead{} 
}
\startdata
HIP490&1.91&0.45&0.0003\\\hline
HIP560&-0.88&0.14&-0.0013\\\hline
HIP1113&5.30&0.98&0.0030\\\hline
HIP1134&0.25&0.30&-0.0014\\\hline
HIP1481&1.36&0.34&-0.0005\\\hline
HIP1910&5.74&1.00&0.0034\\\hline
HIP1993&3.00&0.56&0.0040\\\hline
GJ2006A&6.53&1.46&0.0006\\
\enddata
\end{deluxetable*}
\clearpage

\bibliographystyle{yahapj}

\bibliography{bib}

\end{document}